\documentclass[aps,prd,twocolumn,showpacs,eqsecnum,nofootinbib,amsmath,amssymb,amsfonts,english]{revtex4}

\usepackage{babel}
\usepackage{graphicx}

\begin{document}


\title{Radiation from accelerated black holes in an anti--de~Sitter universe}

\author{Ji\v{r}\'{\i} Podolsk\'y}
\email{Jiri.Podolsky@mff.cuni.cz}

\author{Marcello Ortaggio}
\email{Marcello.Ortaggio@comune.re.it}

\author{Pavel Krtou\v{s}}
\email{Pavel.Krtous@mff.cuni.cz}

\affiliation{
  Institute of Theoretical Physics,
  Faculty of Mathematics and Physics, Charles University in Prague,\\
  V Hole\v{s}ovi\v{c}k\'{a}ch 2, 180 00 Prague 8, Czech Republic
  }

\date{December 1, 2003}     

\begin{abstract}
We study gravitational and electromagnetic radiation generated by uniformly
accelerated charged black holes in anti--de~Sitter spacetime. This
is described by the $C$-metric exact solution of the Einstein-Maxwell
equations with a negative  cosmological constant~$\Lambda$.
We explicitly find and interpret the pattern of radiation that characterizes the
dependence of the fields on a null direction from which the (timelike) conformal
infinity is approached. This directional pattern exhibits specific  properties
which are  more complicated if compared with recent analogous results
obtained for asymptotic behavior of fields near a de~Sitter-like infinity.
In particular, for large acceleration the anti--de~Sitter-like infinity is divided
by Killing horizons into several distinct domains with a different structure
of principal null directions, in which the patterns of radiation differ.

\end{abstract}



\maketitle


\section{Introduction}
\label{sc:intro}

In the context of exact solutions of Einstein's field equations, gravitational radiation
has been studied for decades. In particular, various techniques have been developed
to rigorously characterize  asymptotic properties of the gravitational field, i.e.,
the geometry of spacetime at \vague{large distance} from bounded sources.

In the fundamental work \cite{BondiBurgMetzner:1962} gravitational waves emitted by
axisymmetric systems were  analyzed  by considering an expansion of metric functions
in inverse powers of an appropriate  \vague{radial} coordinate $r$ parametrizing
outgoing null geodesics. In particular, the news function was defined that characterizes
radiation, and which is related to a decreasing (Bondi) mass of the source.
Generalizations and refinement of this method, with a deeper understanding of its relation to
the Petrov types (such as the peeling-off behavior of the Weyl tensor) were subsequently
achieved in \cite{Sachs:1962,NewmanPenrose:1962,NewmanUnti:1962,Burg:1969,BicakPravdova:1998},
see, e.g., \cite{Pirani:1965,Bicak:Bonnor,Bicak:1997,Bicak:Ehlers} for reviews.
Nevertheless, in these works the analysis of radiative fields remained confined to
\emph{asymptotically flat} spacetimes thus ruling out, for instance, the presence of a nonvanishing
cosmological constant $\Lambda$.  In addition, it was based on the use of privileged coordinate systems.

It was Penrose \cite{Penrose:1963,Penrose:1964,Penrose:1965}, see \cite{PenroseRindler:book2}
for a comprehensive overview, who introduced a {\em covariant} approach to the definition of
radiation for massless fields, which is based on the conformal treatment of infinity
(a comparison of the Bondi-Sachs and Penrose approaches was recently presented in \cite{TafelPukas:2000}).
This enables one to apply methods of local differential geometry \vague{at infinity}, and thus
to define in a rigorous  geometric way such basic concepts as the Bondi mass, the
peeling-off property, and the Bondi-Metzner-Sachs group of asymptotic symmetries.
In particular, gravitational radiation propagating along a given null geodesic
is described by the $\WTP{}{4}$ component of the Weyl tensor projected on a parallelly
transported complex null tetrad at infinity. The crucial point is that such a tetrad is
(essentially) determined  {\em uniquely} by the conformal geometry \cite{PenroseRindler:book2}.
Moreover, an advantage of the Penrose method is that it can be naturally
applied also to asymptotically simple spacetimes which \emph{include the cosmological
constant} \cite{Penrose:1964,Penrose:1965,PenroseRindler:book2}. This is quite remarkable since there
is no analogue of the news function in the presence of $\Lambda$
\cite{AshtekarMagnon:1984,AshtekarDas:2000}.

However, specific new features appear in the case of asymptotically de~Sitter
($\Lambda>0$) or anti--de~Sitter ($\Lambda<0$) spacetimes, for which the
conformal infinity~$\scri$ is, respectively, spacelike or timelike
\cite{Penrose:1964,Penrose:1965,PenroseRindler:book2,Penrose:1967}. First of
all, the concept  of radiation for a massless field is \vague{less invariant}
in cases when $\scri$ is not  null. Namely, it emerges as necessarily
direction dependent since the choice of the above-mentioned
null tetrad, and thus the radiative component $\WTP{}{4}$ of the
field, turns out to be \emph{different} for different null geodesics reaching
the same point on $\scri$.
This is related to the fact that with nonvanishing $\Lambda$
even fields of \emph{nonaccelerated} sources are \emph{radiative}
along a \emph{generic} direction,
as it has been shown for test charges \cite{BicakKrtous:2002} or
for Reissner-Nordstr\"om black holes \cite{KrtousPodolsky:2003a} in a de~Sitter universe,
and it will be shown here for a negative $\Lambda$ (Sec.~\ref{ssc:vanishingA}).
In addition, the character of infinity plays a
crucial role in the formulation of the initial value problem.
A spacelike $\scri$ implies the insufficiency of purely retarded massless
fields so that, for example, in de~Sitter space
purely retarded solutions of the Maxwell equations are impossible for
generic charge distributions
\cite{BicakKrtous:2001}. On the other hand, it is wellknown
that a timelike $\scri$ prevents the existence of a Cauchy surface, and
one is necessarily led to a kind of \vague{mixed initial value boundary
problem}, see, e.g., \cite{Avisetal:1978,Hawking:1983,Friedrich:1995}.
For all the above reasons, the definition of radiation is much less obvious when
${\Lambda\not=0}$.

Any explicit exact example of a source which generates
gravitational waves in an (anti--)de~Sitter universe is thus of
paramount importance since this may provide us with insight
into the character of radiation in spacetimes which
are not asymptotically flat. Exact solutions
with boost-rotation symmetry
\cite{BicakSchmidt:1989, Bicak:1968, Pravdovi:2000}, which represent
radiative spacetimes with uniformly accelerating sources,
play a unique role when ${\Lambda=0}$. Among these
the $C$-metric, which describes accelerated black holes,
admits a natural generalization to a nonvanishing value of the cosmological
constant, and it will thus be considered in the present paper.

The $C$-metric \cite{LeviCivita:1917,Weyl:1919,EhlersKundt:1962,Krameretal:book}
is a classic solution of the Einstein(-Maxwell) equations which has
been physically interpreted and analyzed in fundamental papers
\cite{KinnersleyWalker:1970,FarhooshZimmerman:1979,AshtekarDray:1981,Bonnor:1982}
and in many other works, see e.g.
\cite{BicakSchmidt:1989,BicakPravda:1999,Pravdovi:2000,LetelierOliveira:2001,DiasLemos:2003a}
for references and summary of the results. A generalization of the standard
$C$-metric to admit a nonvanishing value of $\Lambda$ has also been known for
a long time \cite{PlebanskiDemianski:1976}, cf.\ \cite{Carter:1968,Debever:1971}
(also, related solutions have been obtained by considering extremal limits
of the $C$-metric with an arbitrary $\Lambda$ \cite{DiasLemos:2003c}).
These spacetimes have found successful application to the
problem of cosmological pair creation of black holes
\cite{MannRoss:1995,Mann:1997,Mann:1998,BoothMann:1999}.
However, a deeper understanding of their physical and global properties,
including the character of radiation, has been missing until recently.
The interpretation of the $C$-metric solutions with ${\Lambda>0}$,
in particular the meaning of parameters in the metric and the relation to
the \vague{background} de~Sitter universe, was clarified in \cite{PodolskyGriffiths:2001}
by introducing an appropriate coordinate system adapted to uniformly accelerated
observers. The causal structure was further studied in \cite{DiasLemos:2003b}
for various choices  of the physical parameters. Very recently
\cite{KrtousPodolsky:2003a}, we have carefully analyzed the $C$-metric with
${\Lambda>0}$ and, among other results, we have demonstrated that
gravitational and electromagnetic fields  of this exact solution exhibit
asymptotically a \emph{specific directional pattern of  radiation} at $\scri$.
Interestingly, this directional dependence of fields on null directions from
which the conformal infinity is approached is the same as for the test fields
of uniformly accelerated charges  in a de~Sitter universe
\cite{BicakKrtous:2002}.

In the present work we wish to investigate an analogous asymptotic behavior of fields
of the $C$-metric with ${\Lambda<0}$, i.e.,  the directional dependence
at conformal infinity $\scri$ of radiation  generated by uniformly accelerated
(possibly charged) black holes in an anti--de~Sitter universe. Some fundamental
differences from the cases ${\Lambda\ge0}$ appear since $\scri$ now has a
\emph{timelike} character. In fact, the whole structure of the \vague{anti--de~Sitter $C$-metric}
is much more complex and new peculiar phenomena thus occur. As observed in \cite{Emparanetal:2000b}
and thoroughly studied in the recent work \cite{DiasLemos:2003a}, for a small value of acceleration,
${\accl<\sqrt{-\Lambda/3}}$, the metric describes a \emph{single} uniformly accelerated black hole
in an anti--de~Sitter universe \cite{Podolsky:2002} whereas for ${\accl>\sqrt{-\Lambda/3}}$
this represents \emph{a pair} of accelerated black holes. The \vague{limiting case} given by
${\accl=\sqrt{-\Lambda/3}}$, previously investigated in
\cite{Emparanetal:2000a,Chamblin:2001},
plays a special important role in the context of the Randall-Sundrum model
since it describes a black hole bound to a two-brane in four dimensions.
However, this case is not investigated in the present work.

Our paper is organized as follows. First, in Sec.~\ref{sc:Cmetric} we
present the $C$-metric solution with a negative cosmological
constant, in particular the Robinson-Trautman coordinates which will be
used in  the subsequent analysis. Basic properties of the solution are
also summarized, including a description of the global structure.
Secs.~\ref{sc:RadChar}--\ref{sc:Analysis} contain  the core of our analysis.
First we define a suitable interpretation tetrad
parallelly transported along null geodesics approaching asymptotically
a given point on conformal infinity $\scri$ from all possible spacetime directions.
The magnitude of the leading terms of gravitational and electromagnetic
fields in such a tetrad then provides us with a specific directional
pattern of radiation. Convenient  parametrizations  of null
directions approaching $\scri$ are introduced in Sec.~\ref{sc:param}, and
the results are subsequently described and analyzed
in Sec.~\ref{sc:Analysis}. This is done for both the cases of a single
black hole and a pair of black holes accelerating in an anti--de~Sitter universe
(and for vanishing acceleration).
The paper also contains two appendixes. In Appendix~\ref{apx:AlgSpecDir}
the behavior of radiation along special null directions is studied.
In particular, for geodesics along principal null directions
the results are obtained in closed explicit
form without performing asymptotic expansions of the physical quantities near ${\scri}$.
Appendix~\ref{apx:Transformations} summarizes the Lorentz
transformations of the null-tetrad components of the
gravitational and electromagnetic fields.

\section{The $C$-metric with a negative cosmological constant}
\label{sc:Cmetric}

The  $C$-metric with a cosmological constant ${\Lambda<0}$, contained in the
family of solutions \cite{PlebanskiDemianski:1976}, can be written as
\begin{equation}\label{KWmetric}
  \mtrc =
  \frac1{\accl^2(\xKW+\yKW)^2}\Bigl(
    -\FKW \,\grad\tKW^2
    +\frac1{\FKW} \,\grad\yKW^2
    +\frac1{\GKW} \,\grad\xKW^2
    +\GKW \,\grad\ph^2
    \Bigr)\commae
\end{equation}
where $\FKW$ and $\GKW$ are, respectively, polynomials of
$\yKW$ and $\xKW$,
\begin{equation}\label{KWFG}
\begin{aligned}
  \FKW &= \frac{-\Lambda}{\ \>3\accl^2}-1+\yKW^2
  -2\mass\accl\,\yKW^3+\charge^2\accl^2\,\yKW^4\commae\\
  \GKW &= \mspace{67mu} 1-\xKW^2
  -2\mass\accl\,\xKW^3-\charge^2\accl^2\,\xKW^4\period
\end{aligned}
\end{equation}
These functions are mutually related by
\begin{equation}\label{ae:KWFGQ}
  \FKW = -\QKW(\yKW)+\frac{-\Lambda}{\ \>3\accl^2}\comma
  \GKW = \QKW(-\xKW)\commae
\end{equation}
where ${\QKW(w)} = 1 - w^2 + 2\,\mass\accl\, w^3 - \charge^2\accl^2\,w^4$.
The metric \eqref{KWmetric}
is a solution of the Einstein-Maxwell
equations with a non-null electromagnetic field given by
\begin{equation}\label{KWEMF}
  \EMF = \charge\, \grad\yKW\wedge\grad\tKW\commae
\end{equation}
or related expressions which can be obtained by a constant duality rotation.
There exist two double-degenerate principal null directions (PNDs)
\begin{equation}\label{PNDsKW}
   \kG_1 \propto\cvil{\tKW}-\FKW\cvil{\yKW}   \comma
   \kG_2 \propto\cvil{\tKW}+\FKW\cvil{\yKW}   \commae
\end{equation}
so that the spacetime is of the Petrov type $D$.
It admits two \defterm{Killing vectors} $\cvil{\tKW}$, $\cvil{\ph}$, and
one \defterm{conformal Killing tensor $\tens{Q}$}
(cf.\ Refs.~\cite{WalkerPenrose:1970,Hughstonetal:1972,PenroseRindler:book2}),
\begin{equation}\label{KillTens}
  \tens{Q} =
  \frac1{\accl^4(\xKW+\yKW)^4}\Bigl(
    \FKW \,\grad\tKW^2
    -\frac1{\FKW} \,\grad\yKW^2
    +\frac1{\GKW} \,\grad\xKW^2
    +\GKW \,\grad\ph^2
    \Bigr) \period
\end{equation}

The metric \eqref{KWmetric} can describe different spacetimes,
depending on the choice of parameters and of ranges of coordinates.
We are interested in the physically most relevant case when the metric
describes one black hole or pairs of black holes
uniformly accelerated in anti--de~Sitter universe.
In this case the constants $\accl$, $\mass$, $\charge$, and $\conpar$,
such that ${\ph\in(-\pi\conpar,\pi\conpar)}$,
characterize acceleration, mass, charge of the black holes, and conicity
of the $\ph$ symmetry axis, respectively.
They have to satisfy $\mass\ge0$, $\charge^2<\mass^2$, ${\accl,\,C>0}$,
and they have to be such that the function $\GKW$ has
four real roots in the charged case (${\charge,\,\mass\neq0}$) or
three real roots in the uncharged case (${\charge=0}$, ${\mass\neq0}$).
The coordinates ${\xKW,\,\yKW}$ have to satisfy ${\yKW>-\xKW}$
and ${\xKW\in[\xf,\xb]}$, where ${\xf<0<\xb}$ are two roots of $\GKW$,
namely those closest to zero
(see \cite{KrtousPodolsky:2003a,Mann:1997,Podolsky:2002,DiasLemos:2003a} for details and
discussion of other cases, cf.\ also Figs.~\ref{fig:CAdSI}(d) and \ref{fig:CAdSIIxy} below).
From these conditions we obtain $0\le\GKW\le1$, and $\mass+2\charge^2\accl\,\xKW>0$.
The spacetime \eqref{KWmetric} reduces to the anti--de~Sitter universe for ${\mass=0,\,\charge=0}$.

The coordinates $\xKW$ and $\ph$ are longitudinal and latitudinal angular coordinates,
$\xf$ denotes the axis of $\ph$ symmetry extending from the \vague{forward} pole
of the black hole (in the direction of acceleration) to infinity,
whereas $\xb$ denotes the axis from the opposite \vague{backward} pole.
For nonvanishing acceleration the axis cannot be regular everywhere --- at least one
part of it has to have a nontrivial conicity, depending on the choice
of the parameter $\conpar$. This corresponds to the
presence of cosmic strings (or struts) which are responsible for the
acceleration of the black holes, see the references above for more details.

The spacetime metric \eqref{KWmetric} can be put into various forms.
In this paper we concentrate on investigation of radiation near infinity,
for which the Robinson-Trautman
form seems to be a convenient one. Introducing real coordinates
$\rRT$, $\uRT$ and complex coordinates ${\zRT,\,\bRT}$ by
\begin{equation}\label{RTKW}
\begin{aligned}
  \accl\,\rRT &= (\xKW+\yKW)^{-1}\commae\\
  \accl\,\grad\uRT &= \frac{\grad\yKW}{\FKW}+\grad\tKW\commae\\
  \textstyle{\frac{1}{\sqrt2}}  (\grad\zRT+\grad\bRT)&= \frac{\grad\yKW}{\FKW}-\frac{\grad\xKW}{\GKW}+\grad\tKW\commae\\
  \textstyle{\frac{i}{\sqrt2}}(\grad\zRT-\grad\bRT)&= \grad\ph  \commae
\end{aligned}
\end{equation}
we put the $C$-metric \eqref{KWmetric} into the form\footnote{%
The symmetric product $\stp$ of two 1-forms is defined as
${\tens{a}\stp\tens{b}}={\tens{a}\,\tens{b}+\tens{b}\,\tens{a}}$.}
\begin{equation}\label{RTmetric}
  \mtrc =
  \frac{\rRT^2}{\PRT^2}\,\grad\zRT\stp\grad\bRT
    -\grad \uRT\stp\grad\rRT-\HRT\,\grad\uRT^2 \period
\end{equation}
The metric functions are
\begin{equation}\label{ae:PandH}
  \PRT^{-2} = \GKW     \commae \qquad\
  \HRT  =\accl^2\rRT^2(\FKW+\GKW) \commae
\end{equation}
or explicitly
\begin{equation}\label{ae:cosi}
\begin{aligned}
 \HRT &=-\frac{\Lambda}{3}\,\rRT^2-2\,\rRT\,(\ln\PRT)_{,u}+\Delta\ln\PRT\\
 &\qquad\qquad\quad -\frac{2}{\rRT}(\mass+2\charge^2\accl\,\xKW)+
 \frac{\charge^2}{\rRT^2}\commae
\end{aligned}
\end{equation}
with ${\Delta=2\PRT^2\partial_{\zRT}\partial_{\bRT}}$,
where $\xKW$ is expressed using the relation
\begin{equation}\label{ae:relforx}
  \int \frac{d\xKW }{\GKW(\xKW)} = \accl\,\uRT-
    \textstyle{\frac{1}{\sqrt2}}  (\zRT+\bRT)   \period
\end{equation}
The functions \eqref{ae:PandH}, \eqref{ae:cosi} represent a particular case, corresponding to the $C$-metric,
of the standard general expression for the Robinson-Trautman spacetimes \cite{Krameretal:book}.
As opposed to the metric form \eqref{KWmetric}, the
Robinson-Trautman coordinates allow an explicit limit ${\accl=0}$.
The coordinates are not defined globally but
it is possible to cover the whole universe by many coordinate
patches of the same type. Therefore, it is sufficient to study the
spacetime only in just one Robinson-Trautman coordinate map;
such a patch is indicated by a shaded domain in
Figs.~\ref{fig:CAdSIIA}--\ref{fig:CAdSIIC}.
We additionally assume there that the coordinate $\uRT$ is
increasing from the past to the future.

The global causal structure of the $C$-metric with ${\Lambda<0}$ has recently been
analyzed in \cite{DiasLemos:2003a}. In particular,
the character of infinity, singularities, and possible horizons has been
described in detail.

\emph{Infinity} $\scri$ of the spacetime is given by
\begin{equation}\label{infinity}
  \rRT=\infty\comma\text{or equivalently}\quad  \xKW+\yKW=0 \commae
\end{equation}
where the conformal factor
\begin{equation}
 \Omega=\frac{1}{\rRT}=\accl(\xKW+\yKW)
\end{equation}
vanishes. The conformal (unphysical) metric
${\bf\tilde{\mtrc}}=\Omega^2\mtrc\commae$
\begin{equation}\label{confmetric}
  \cmtrc =
  \frac{1}{\PRT^2}\,\grad\zRT\stp\grad\bRT
    +\grad \uRT\stp\grad\Omega-\HRT\,\Omega^2\grad\uRT^2 \period
\end{equation}
is regular at infinity, given by $\Omega=0$.
Moreover, it follows from Eq.~\eqref{ae:cosi} that at $\scri$ the metric function reads
\begin{equation}\label{Hasympto}
\HRT\,\Omega^2\Big|_{\scri}= -\frac{\Lambda}{3} \commae
\end{equation}
i.e.. it is independent of the parameters $\mass$, $\charge$, and $\accl$.
The vector ${\Norm=-(\HRT\Omega^2\cv{\Omega}+\cv{u})}$ is orthogonal to each
hypersurface $\Omega=\rm{constant}$. In particular, it is outgoing and normal
to infinity $\scri$ at any of its point, with the norm
${{\bf\tilde{\mtrc}}(\Norm,\Norm)=\HRT\Omega^2=-\frac{\Lambda}{3}>0}$.
The universe is thus weakly asymptotically anti--de~Sitter \cite{AshtekarMagnon:1984},
at least locally, with the conformal infinity $\scri$ having a timelike
character (in general, it is not asymptotically anti--de~Sitter
according to the definition based on the \vague{reflective boundary condition}
\cite{Hawking:1983,AshtekarMagnon:1984,HenneauxTeitelboim:1985,AshtekarDas:2000}:
the ${(2+1)}$-metric induced on $\scri$ by ${\bf\tilde{\mtrc}}$ is not
conformally flat since  the associated Bach tensor is nonvanishing).
Throughout the paper, however, it will be more convenient to employ the spacelike
outward vector ${\norm=\Norm/\sqrt{\HRT}}$ orthogonal to $\scri$,
\begin{equation}\label{gennorm}
\norm=\sqrt{\HRT}\,\cv{\rRT}-\frac{1}{\sqrt{\HRT}}\,\cv{\uRT}  \commae
\end{equation}
which has a unit norm
${\norm\spr\norm =\mtrc(\norm,\norm)= 1}$ with respect to the \emph{physical} metric.

At ${\rRT=0}$ the metric has unbounded curvature which corresponds to
a physical singularity hidden behind the black hole horizon.
Similarly to the $C$-metric with vanishing
$\Lambda$ \cite{KinnersleyWalker:1970,FarhooshZimmerman:1979,AshtekarDray:1981,Bonnor:1982,BicakPravda:1999,LetelierOliveira:2001}
or the $C$-metric with ${\Lambda>0}$ \cite{PodolskyGriffiths:2001,DiasLemos:2003b,KrtousPodolsky:2003a},
the zeros of the function $\FKW$ correspond to  \emph{Killing horizons} associated with $\cvil{\tKW}$.
Interestingly, unlike in the ${\Lambda\ge0}$ case,
the anti--de~Sitter $C$-metric describes either a single uniformly accelerated black hole
(for ${\accl<\sqrt{-\Lambda/3}}$~), or a pair of uniformly accelerated black holes
(when ${\accl>\sqrt{-\Lambda/3}}$~).

\subsection{A single accelerated black hole}
\label{ssc:single}

\begin{figure}
\vspace*{0.8em}
\includegraphics{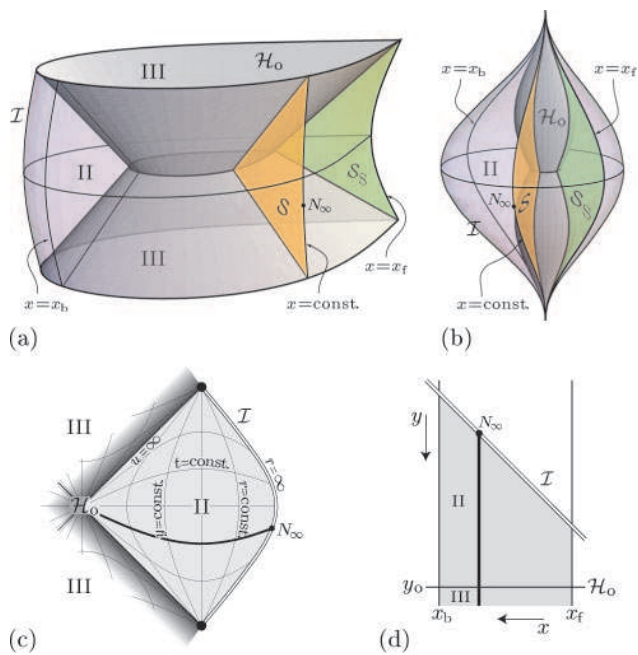}
\caption{\label{fig:CAdSI}
Spacetime outside a single accelerated black hole
moving in an anti--de~Sitter universe
with acceleration ${\accl<\sqrt{-\Lambda/3}}$.
Diagrams~(a) and (b) depict a three-dimensional section
${\ph=\text{constant}}$, diagram~(c) is a conformal diagram
of the ${\tKW\textdash\yKW}$ section, and diagram~(d) shows
relevant ranges of coordinates $\xKW$ and $\yKW$.
The diagrams focus on region~II between the outer black hole
horizon $\horo$ (${\yKW=\yo}$) and timelike infinity $\scri$, ${\yKW=-\xKW}$.
Therefore, only a small part of an interior of the black hole
near the horizon $\horo$ is shown (region III).
In diagram~(a) the horizon $\horo$ is represented by two conical surfaces
which intersect on a bifurcation surface of the Killing vector $\cvil{\tKW}$
(a continuation of cones inside the black hole to another asymptotic
domain is not shown). The outer deformed boundary of domain II
corresponds to infinity $\scri$. A particular section $\sect{}$
given by ${\xKW=\text{constant}}$ is shown, and the axes
${\xKW=\xb}$ and ${\xKW=\xf}$ are indicated.
It is assumed in diagram~(a) that the string causing
acceleration of the black hole is located on the
\vague{forward} axis ${\xKW=\xf}$ and the corresponding conical
singularity is represented by nonsmooth behavior at $\sect{S}$
(the edge at ${\xKW=\xf}$).
Diagram~(b) is a deformation of diagram~(a)
in which both the top and the bottom of the diagram are
squeezed to single points, and the longitudinal
$\xKW$~direction is embedded smoothly at both axes.
The black hole horizon $\horo$ has thus a droplike shape,
symmetrical around the vertical axis.
Diagram~(c) is the Penrose-Carter conformal diagram of
the ${\tKW\textdash\yKW}$ section $\sect{}$ (${\xKW,\ph=\text{constant}}$).
Both principal null directions lie in this section.
The precise shape of infinity $\scri$ (double line) depends on the value
of coordinate $\xKW$, cf.\ Eq.~\eqref{infinity} [and this
dependence is the reason for the deformed shape of $\scri$ in diagram~(a)].
Lines ${\tKW=\text{constant}}$, and ${\yKW=\text{constant}}$
(coinciding with ${\rRT=\text{constant}}$ in $\sect{}$) are drawn
with labels oriented in the direction of an increasing coordinate. A small part of
the interior of the black hole is indicated by the dark area at the left of the diagram.
Finally, diagram~(d) depicts the ${\xKW\textdash\yKW}$ section
for relevant ranges of coordinates (shaded area). The infinity is again
represented by the double line, and the horizon $\horo$ is shown.
The thick line ${\xKW=\text{constant}}$ corresponds to the
${\tKW\textdash\yKW}$ section of diagram~(c),
similarly the thick line ${\tKW=\text{constant}}$
in diagram~(c) corresponds to the ${\xKW\textdash\yKW}$ section
from diagram~(d).}
\end{figure}

Indeed, as described in \cite{Podolsky:2002,DiasLemos:2003a}, when the acceleration parameter
$\accl$ is \emph{small}, namely ${\accl<\sqrt{-\Lambda/3}}$,
and $\mass\neq0$, the metric
\eqref{KWmetric}, \eqref{KWFG} describes a \emph{single} uniformly accelerated
black hole. The condition of small acceleration guarantees that
the function $\FKW$ has only two zeros $\yo,\,\yi$ in the charged case,
or only one zero $\yo$ for the uncharged black hole.
These zeros define \defterm{outer} and, if applicable, \defterm{inner} horizons of the black hole.
The relevant ranges of coordinates $\xKW,\,\yKW$ representing the spacetime
outside the black hole are depicted in Fig.~\ref{fig:CAdSI}(d).

In \cite{DiasLemos:2003a} the causal structure
of this spacetime was represented by the Penrose-Carter
conformal diagram of a two-dimensional ${\tKW\textdash\yKW}$ section
(i.e., the section of constant angular coordinates ${\xKW,\,\ph}$).
This section is, in fact, spanned by the PNDs $\kG_1$ and  $\kG_2$, cf.\ Eq.~\eqref{PNDsKW}.
A part of such a conformal diagram representing
an exterior of the black hole is depicted in Fig.~\ref{fig:CAdSI}(c).
The conformal infinity  $\scri$ is indicated here by a double line.
The outer horizons $\horo$, given by ${\yKW=\yo}$, separate
region II outside the black hole from an interior of the black
hole, denoted as III%
.\footnote{There is no region I in this case. The numbering of regions, starting
  with II, has been chosen to be consistent with the case of two black holes (cf.\
  Figs.~\ref{fig:CAdSIIA}--\ref{fig:CAdSIIC}) and with the $C$-metric with
  $\Lambda>0$ \cite{KrtousPodolsky:2003a}.}
A more detailed structure of the interior of the black hole
depends on whether the hole is charged or not, and its causal structure is analogous
to the Schwarzschild or Reissner-Nordstr\"om black holes. Because we are mainly
interested in region II near the infinity, we will not discuss the interior of the hole here.

It seems to be more instructive to combine the ${\tKW\textdash\yKW}$ sections
for different values of $\xKW$ into a unifying three-dimensional
picture in which just the coordinate $\ph$ is suppressed,
as it is done in Fig.~\ref{fig:CAdSI}(a). Despite
the fact that this is not a complete and rigorous conformal diagram,
it helps to visualize and understand the global causal structure of the spacetime.
The outer horizon $\horo$ of the black hole is here indicated by two joined
conical surfaces, and the conformal timelike infinity $\scri$ is depicted
as a deformed outer boundary.
It has a \vague{simple} topology ${\realn\times\mathrm{S}^2}$ if we include
\vague{nonsmooth} points on the $\ph$ axis where the string is located.
For a vanishing acceleration the timelike infinity would be
rotationally symmetric around the vertical axis, and smooth everywhere.
Its deformation for ${\accl\neq0}$
indicates that the coordinates are adapted to the accelerated source
(for an analogous discussion in the ${\Lambda>0}$ case see \cite{KrtousPodolsky:2003a})
and that there is a string (a conical singularity) on the $\ph$~axis.
Particular surfaces $\sect{}$ of a constant $\xKW$, corresponding to the two-dimensional
conformal diagram~\ref{fig:CAdSI}(c), are also indicated.
The section $\sect{S}$ with ${\xKW=\xf}$ corresponds
to the axis from the \vague{forward} pole of the black hole,
the value ${\xKW=\xb}$ to the axis from the opposite \vague{backward} pole.
In Fig.~\ref{fig:CAdSI}(a) the conicity parameter $\conpar$ is chosen in such a way that
the string is located only on the axis ${\xKW=\xf}$,
and the conical singularity is indicated by nonsmooth embedding of the
${\tKW,\,\yKW=\text{constant}}$ surface into the three-dimensional diagram,
i.e., by a nonsmooth gluing of ${\xKW=\text{constant}}$ sections at $\xf$.
Notice that since ${\FKW>0}$ for $\yKW<\yo$,
region~II near infinity $\scri$ is everywhere
\emph{static}, and there are no Killing horizons which extend up to $\scri$
[cf.\ also Eqs.~\eqref{FtoSource}, \eqref{noKill}].
We can thus interpret the spacetime as a universe having \emph{global}
anti--de~Sitter-like infinity (except the nonsmoothness at the string)
with the black hole moving \vague{inside} it (in contrast to the case discussed below,
where pairs of black holes \vague{enter} and \vague{exit} the spacetime \vague{through} the infinity).

The diagram~\ref{fig:CAdSI}(b) is a deformed version of the diagram~\ref{fig:CAdSI}(a):
gluing of the angular coordinate $\xKW$ is now done smoothly even on the axis
where the string is located, and the top and the bottom of the diagram
are \vague{squeezed} to single points. The horizon $\horo$ thus has
a shape of two joined \vague{drops}, and sections ${\xKW=\text{constant}}$
are deformed accordingly.
Here we can see that
coordinates $\tKW,\,\yKW,\,\xKW,\,\ph$ used to construct this diagram
are adapted to the source, not to the infinity ---
the horizon $\horo$ is symmetric around the vertical axis in contrast to
infinity  $\scri$ which is deformed in the direction of acceleration.
Diagram~\ref{fig:CAdSI}(b) is not so crucial in the present case of a single
black hole but an analogous representation of the black hole horizon will be used in
the case of two accelerated holes which we are going to discuss now.

\subsection{A pair of accelerated black holes}
\label{ssc:pair}

\begin{figure}
\includegraphics{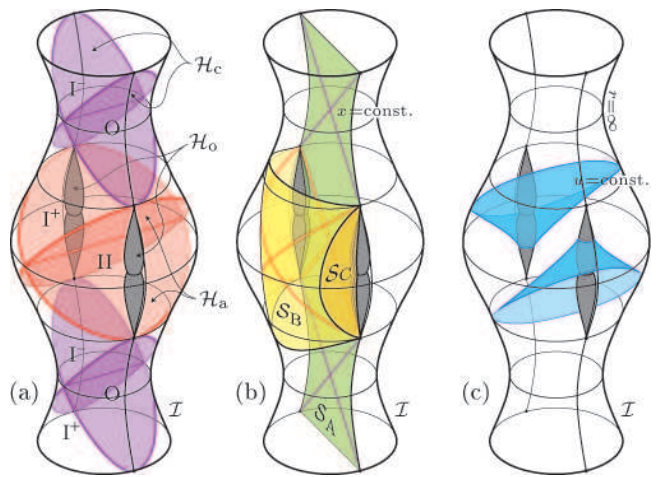}
\caption{\label{fig:CAdSII}
Schematic diagrams of a part of spacetime outside accelerated black holes
moving in an anti--de~Sitter universe
with acceleration ${\accl>\sqrt{-\Lambda/3}}$.
Diagrams represent a three-dimensional section ${\ph=\text{constant}}$.
They depict a domain near one pair of black holes. However,
they should continue periodically in a vertical direction, featuring
thus an infinite chain of pairs of black holes entering and later exiting the spacetime
through timelike infinity $\scri$.
Only regions of spacetime \emph{outside} the outer black hole horizons $\horo$
are drawn. Interiors of black holes and continuations of the spacetime into
other asymptotically anti--de~Sitter universes (through the Einstein-Rosen bridge or
through charged black hole) are hidden under the horizons $\horo$
and not studied in the paper.
The outer black hole horizons $\horo$ are represented by droplike gray shapes
analogous to that of Fig.~\ref{fig:CAdSI}(b), the timelike infinity $\scri$
is depicted as a deformed cylindrical boundary of the diagrams.
Diagram (a) shows the Killing horizons outside the black holes:
\defterm{acceleration horizons} $\hora$ separating two black holes,
and \defterm{cosmological horizons} $\horc$ separating different pairs of black holes
(only one pair of holes is drawn in the diagram). These horizons divide spacetime into
several regions: static domains O and II, nonstatic domains $\Iplus$ and $\Iminus$,
and domains inside the holes hidden under $\horo$.
Diagram (b) indicates embedding of ${\tKW\textdash\yKW}$ sections for different constant values
of the coordinate $\xKW$, which are spanned by principal null directions.
Three qualitatively different sections $\sect{A}$, $\sect{B}$, and
$\sect{C}$ are shown. Exact conformal diagrams corresponding to
these sections can be found in Figs.~\ref{fig:CAdSIIA}--\ref{fig:CAdSIIC}.
Diagram (c) shows hypersurfaces ${\uRT=\text{constant}}$ which are generated
by null geodesics along the principal null directions $\kG_1$
that are discussed in Appendix~\ref{apx:AlgSpecDir}.
}%
\end{figure}

\begin{figure}
\includegraphics{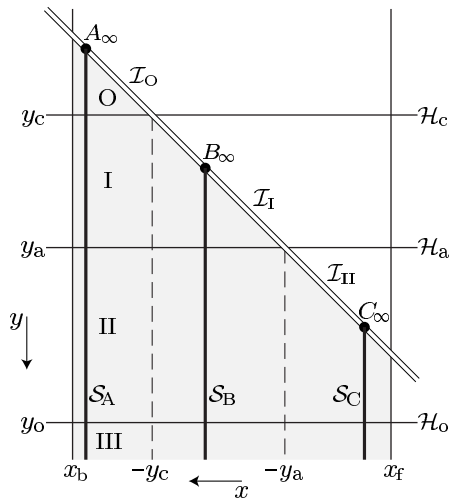}
\caption{\label{fig:CAdSIIxy}
A qualitative ${\xKW\textdash\yKW}$ diagram in the case of large acceleration
${\accl>\sqrt{-\Lambda/3}}$
(cf.\ Fig.~\ref{fig:CAdSI}(d) for small acceleration).
The relevant domain of coordinates $\xKW$ and $\yKW$ is indicated by the shaded area.
It is given by the coordinate $\xKW$ between the axes $\xf$ and $\xb$,
and the coordinate $\yKW$ between the infinity $\scri$, ${\yKW=-\xKW}$, and
the outer black hole horizon $\horo$, ${\yKW=\yo}$
(the interior of black holes, ${\yKW>\yo}$, is not studied here).
The infinity is represented by a double line, the horizons by horizontal lines.
The ${\tKW\textdash\yKW}$ sections of constant $\xKW$ are represented by
vertical lines. Three such typical sections $\sect{A}$, $\sect{B}$ and $\sect{C}$
are shown. They are distinguished by the number of horizons which they intersect. These sections correspond to the
conformal diagrams in Figs.~\ref{fig:CAdSIIA}--\ref{fig:CAdSIIC}.
${A_\onscri,B_\onscri,C_\onscri}$ are points at the infinity which belong to these three sections,
respectively. They can be found also in Figs.~\ref{fig:CAdSIIA}--\ref{fig:CAdSIIC}.}%
\end{figure}

A more complicated situation occurs when ${\accl>\sqrt{-\Lambda/3}}$, ${\mass\neq0}$.
For such large values of acceleration the metric \eqref{KWmetric}, \eqref{KWFG}
describes an infinite number of \emph{pairs} of accelerated black holes
in anti--de~Sitter universe.
In Fig.~\ref{fig:CAdSII}, representing a part of the section ${\ph=\text{constant}}$,
one pair of black holes is indicated by the (outer) horizons $\horo$ which have
droplike shapes analogous to the horizon in diagram~\ref{fig:CAdSI}(b).
The main difference from the previous case is that both
black holes \vague{simultaneously enter} the universe at infinity $\scri$, approach each other
with an opposite deceleration until they stop, and start to move apart, again up to the infinity $\scri$.
The same situation repeats infinitely many times
both in the past and in the future --- the diagrams in Fig.~\ref{fig:CAdSII} should be
infinitely long, composed of parts isomorphic to the part shown there.
In the following we study only one such part of the whole universe.
Relevant ranges of coordinates $\xKW$ and $\yKW$ are drawn in Fig.~\ref{fig:CAdSIIxy}.

\begin{figure}
\includegraphics{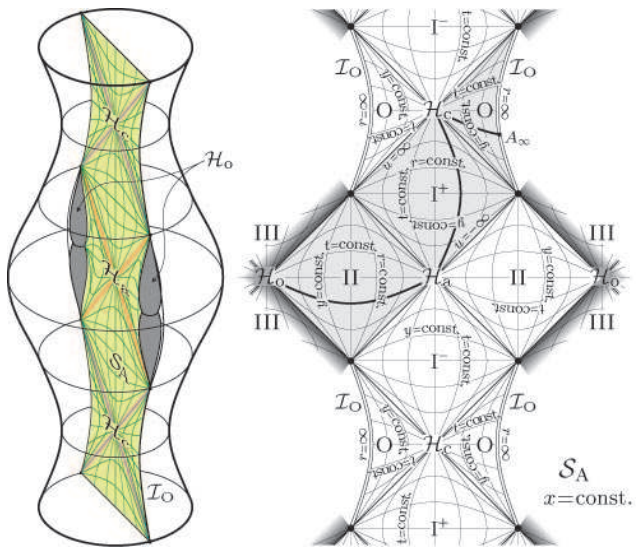}
\caption{\label{fig:CAdSIIA}
Conformal diagram of the ${\tKW\textdash\yKW}$ section $\sect{A}$
intersecting both the horizons $\hora$, $\horc$
outside black holes (right), and its embedding into three-dimensional
diagram ${\ph=\text{constant}}$ (left).
Each of sections of constant $\xKW$ (and $\ph$) with
${\xKW\in(-\yc,\xb]}$ intersects all horizons
and extends through the whole spacetime. Both three-dimensional and two-dimensional
diagrams should continue periodically in the vertical direction.
In the two-dimensional conformal diagram the infinity $\scri$
is depicted by double lines. Section $\sect{A}$ intersects
the infinity in a timelike surface belonging to domain $\scriO$, cf.\ Fig.~\ref{fig:CAdSIIscri}.
Cosmological horizons $\horc$, acceleration horizons $\hora$, and
outer black hole horizons $\horo$ are represented by diagonal lines.
Domains between the horizons are labeled as O, $\Ipm$, II, and III,
cf.\ Fig.~\ref{fig:CAdSII}(a).
Interiors of black holes are indicated only partially, by the dark area
behind the horizon $\horo$. Lines of coordinates $\tKW$ and $\yKW$ are
shown with labels oriented in direction of an increasing coordinate.
An area covered by one Robinson-Trautman coordinate map
(coordinates $\uRT$, $\rRT$) is indicated by the shaded background.
Without the loss of a typical behavior, the special section ${\xKW=\xb}$
has been chosen for this diagram;
other sections with ${\xKW>-\yc}$ look qualitatively the same,
only with embedding not lying in the plane of symmetry
of the three-dimensional diagram.}%
\end{figure}
\begin{figure}
\includegraphics{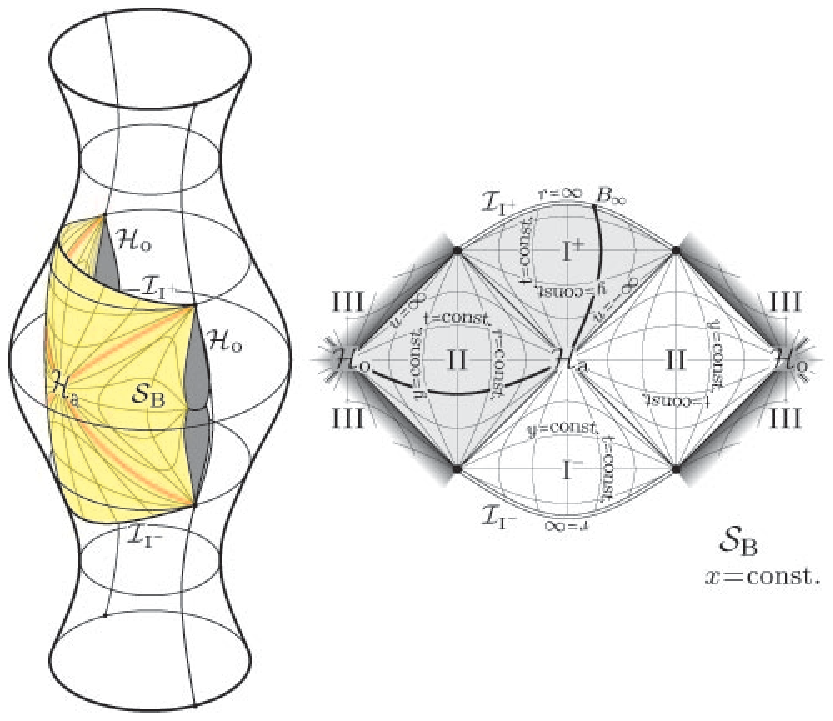}
\caption{\label{fig:CAdSIIB}
Conformal diagram of the ${\tKW\textdash\yKW}$ section $\sect{B}$
intersecting only the horizons $\hora$ outside black holes, and its
embedding into three-dimensional diagram ${\ph=\text{constant}}$.
Such a section of constant $\xKW$ (and $\ph$) with
${\xKW\in(-\ya,-\yc)}$ intersects all horizons except the cosmological
ones. In contrast to section $\sect{A}$ of Fig.~\ref{fig:CAdSIIA},
section $\sect{B}$ does not extend through the whole spacetime, but
it can be found near all pairs of black holes.
The section still extends between both holes of a given pair.
Section $\sect{B}$ intersects
the infinity in two \emph{spacelike} surfaces, one forming a future
boundary of $\sect{B}$ belonging to domain $\scriIp$, the other forming a past boundary
belonging to domain $\scriIm$, cf.\ Fig.~\ref{fig:CAdSIIscri}.
Notation for infinity, horizons, etc., is the same as in
Fig.~\ref{fig:CAdSIIA}. The shaded area again indicates
Robinson-Trautman coordinate patch.}%
\end{figure}

Clearly, both the global causal structure of the spacetime
and the algebraic structure are now more complex.
The metric function $\FKW$ has two more zeros, $\ya$ and $\yc$,
which correspond to the two additional Killing horizons \emph{outside} of
the black holes. We shall refer to these as \defterm{acceleration}
horizons  $\hora$,  and \defterm{cosmological} horizons $\horc$%
.\footnote{The terms \emph{acceleration} and \emph{cosmological} horizons
  are rather arbitrary. Both horizons ${\yKW=\yKW_\ahor}$
  and ${\yKW=\yKW_\chor}$ could actually qualify for the name acceleration
  (Rindler) horizon, cf.\ \cite{DiasLemos:2003a}.
  For convenience, to distinguish them
  we use the name \emph{cosmological} horizon for the horizon ${\yKW=\yKW_\chor}$
  which separates domains containing different pairs of black holes,
  cf.\ Fig.~\ref{fig:CAdSII}(a).}
In contrast to the black hole horizons, spatial sections of these horizons are noncompact.
The horizons  are represented by inclined planes in Fig.~\ref{fig:CAdSII}(a)
or as corresponding horizontal lines in Fig.~\ref{fig:CAdSIIxy}.
They separate static and nonstatic
regions of the spacetime outside of the black holes:
the domains O and II are static, whereas
the domains $\Iplus$ and $\Iminus$ are nonstatic.
Regions II enclose the black holes, the regions O are
\vague{as far as possible away} from the black holes.
Two black holes of a given pair are separated by the acceleration horizons $\hora$,
and they are thus causally disconnected. Different pairs of black holes are
separated by the cosmological horizons $\horc$.

Fig.~\ref{fig:CAdSII}(b) shows the foliation of the spacetime
by the ${\tKW\textdash\yKW}$ surfaces ${\xKW=\text{constant}}$,
the surfaces spanned by the PNDs.
These are of three different types, namely $\sect{A}$, $\sect{B}$, and $\sect{C}$.
Classification of these types is seen from the ${\xKW\textdash\yKW}$
diagram in Fig.~\ref{fig:CAdSIIxy}, where the ${\tKW\textdash\yKW}$ sections
are represented by vertical lines. These different types of ${\tKW\textdash\yKW}$ sections
are distinguished by the number of horizons which they intersect.

For ${\xKW\in(-\yc,\xb]}$ the section denoted as
$\sect{A}$ passes through \emph{all} regions O, I, II (and regions inside black holes),
and intersects all horizons $\horc$, $\hora$, $\horo$. Such a section corresponds to the
conformal diagram drawn in Fig.~\ref{fig:CAdSIIA}. The \emph{intersection}
of such a section with the conformal infinity $\scri$ is \emph{timelike}.
We denote as $\scriO$ a part of the infinity with ${\xKW\in(-\yc,\xb]}$,
i.e., the part which can be reached by the sections $\sect{A}$,
cf.\ also Fig.~\ref{fig:CAdSIIscri}.

The situation is different for
section $\sect{B}$ of a constant ${\xKW\in(-\ya,-\yc)}$ which
goes only through regions I, II (and regions inside black holes),
and does not intersect the horizon $\horc$.
This section corresponds to the conformal diagram presented in Fig.~\ref{fig:CAdSIIB}.
The \emph{intersection} of such a section with infinity is
\emph{spacelike}, and consists of two disjoint parts, one in the future
and another in the past of the section. A part of the infinity
which can be reached in this way will be labeled as $\scriIp$ and $\scriIm$,
respectively, cf.\ Fig.~\ref{fig:CAdSIIscri}.

\begin{figure}
\includegraphics{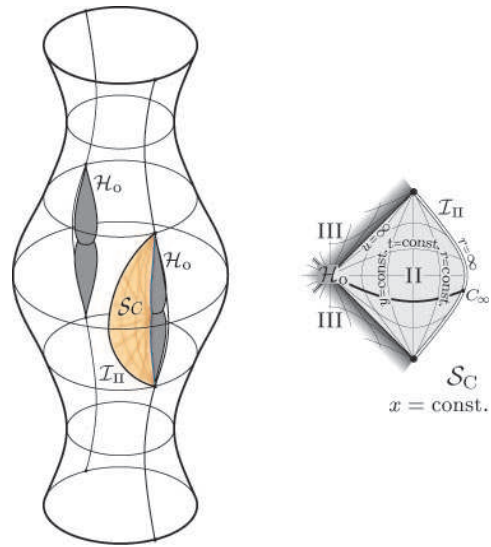}
\caption{\label{fig:CAdSIIC}
Conformal diagram of the ${\tKW\textdash\yKW}$ section $\sect{C}$
not intersecting any horizons outside black holes, and its
embedding into the three-dimensional diagram ${\ph=\text{constant}}$.
Such a section of constant $\xKW$ (and $\ph$) with
${\xKW\in[\xf,-\ya)}$ does not intersect cosmological and acceleration horizons.
In contrast to section $\sect{A}$ of Fig.~\ref{fig:CAdSIIA},
section $\sect{C}$ does not extend through the whole spacetime, but
it can be found near any black hole.
Unlike section $\sect{B}$ from Fig.~\ref{fig:CAdSIIB},
it even does not extend between holes of a given pair of black holes.
Section $\sect{C}$ intersects
the infinity in a timelike surface which belongs to domain $\scriII$
of the conformal infinity, cf.\ Fig.~\ref{fig:CAdSIIscri}.
Notation for infinity, horizons, etc., and the meaning of the shaded area are the same as in
Fig.~\ref{fig:CAdSIIA}.}%
\end{figure}

Finally, for ${\xKW\in[\xf,-\ya)}$ the section $\sect{C}$ of constant $\xKW$
extends only to the region II (and regions inside the black hole),
and it does not intersect the horizons $\hora$ and $\horc$.
The conformal diagram of this type is given in Fig.~\ref{fig:CAdSIIC}.
Section $\sect{C}$ \emph{intersects} the infinity
in a \emph{timelike} surface, and a part of the infinity
which can be reached by these sections will be denoted by $\scriII$,
cf.\ also Fig.~\ref{fig:CAdSIIscri}.

The above described three types of conformal diagrams depicted in
Figs.~\ref{fig:CAdSIIA}--\ref{fig:CAdSIIC} have been drawn recently in
\cite{DiasLemos:2003a} (together with special limiting cases ${\xKW=-\yc}$
and ${\xKW=-\ya}$ which we do not discuss here). Their qualitative dependence
on the value of coordinate $\xKW$ has been already noted there but not discussed in more details.
Putting all these conformal diagrams together to the single three-dimensional
picture shown in Fig.~\ref{fig:CAdSII}(b) elucidates the character of this dependence.
Moreover, it clarifies how the \emph{timelike} infinity $\scri$ can
form a \emph{spacelike} boundary of a conformal diagram
as for section $\sect{B}$, Fig.~\ref{fig:CAdSIIB}
--- this \emph{spacelike} boundary is the \emph{intersection} of
the two \emph{timelike} hypersurfaces $\scri$ and $\sect{B}$.

In Fig.~\ref{fig:CAdSII}(c) the surfaces
${u=\text{constant}}$ are shown. These null surfaces
are formed by null geodesics ${\uRT,\,\zRT=\text{constant}}$
tangent to PND (see Appendix~\ref{apx:AlgSpecDir}), and they indicate
how the spacetime is covered by Robinson-Trautman coordinates.
A surface of constant $\uRT$ reduces to a horizon $\hora$ for
${\uRT=-\infty}$, and to a horizon $\horo$ and $\horc$ for
${\uRT=\infty}$. A connected domain covered by finite values of
coordinate $\uRT$ is indicated in Figs.~\ref{fig:CAdSIIA}--\ref{fig:CAdSIIC}
by the shaded background. Remaining parts of spacetime have to be
covered by different patches of Robinson-Trautman coordinates defined analogously.
The domain indicated in figures by a shaded background,
covered by a single Robinson-Trautman map,
thus reaches up to all types of infinity except to the part $\scriIm$, which is, however,
related to the part $\scriIp$ by a simple time reversion.
Therefore, we do not loose any substantial information using
only this Robinson-Trautman coordinate map.

\section{Gravitational and electromagnetic fields near $\scri$}
\label{sc:RadChar}

Now we are prepared to discuss radiative properties of the $C$-metric fields
near the timelike infinity $\scri$. As we have
already mentioned in Sec.~\ref{sc:intro}, following \cite{PenroseRindler:book2},
by \emph{radiative field} we
understand a field with the dominant component having the $1/\afp$ fall-off,
calculated in a tetrad parallelly transported along a null geodesic $\geod{}(\afp)$,
$\afp$ being the affine parameter. In the following we derive the characteristic
\defterm{directional pattern of radiation}, i.e., the dependence of the radiative component
of the fields on the \emph{direction} along which a \emph{given} point
at the infinity is approached.

We start with a general \emph{null geodesics $\geod(\afp)$}
approaching a fixed point $N_\onscri$ at infinity $\scri$ as ${\afp\to+\infty}$
(or ${\afp\to-\infty}$).
We observe that coordinates ${\xKW,\,\yKW}$, and ${\uRT,\,\zRT}$,
as well as metric functions ${\FKW,\,\GKW}$, and $\PRT$,
are finite at the point $N_\onscri$, whereas the affine parameter $\afp$
and the Robinson-Trautman coordinate $\rRT$ go to infinity
as the geodesic approaches $\scri$, cf.\ Sec.~\ref{sc:Cmetric}.
We denote the limiting values on $\scri$
of the coordinates and the metric functions by a subscript ``$\onscri$''.

To obtain a more detailed description how $N_\onscri$ is reached,
we have to find the tangent vector of the geodesic,
\begin{equation}
\frac{D\geod}{d\afp}=\dot{\tKW}\,\cv{\tKW}+\dot{\yKW}\,\cv{\yKW}
+\dot{\xKW}\,\cv{\xKW}+\dot{\ph}\,\cv{\ph}\period
\end{equation}
As mentioned in \cite{KinnersleyWalker:1970}, an explicit form of
the tangent vector can be obtained using specific geometrical
properties of the $C$-metric (\ref{KWmetric}).
The presence of two Killing vectors $\cv{\tKW}$, $\cv{\ph}$,
and of one conformal Killing tensor $\tens{Q}$ (cf.\ Sec.~\ref{sc:Cmetric})
implies that there exist three constants of motion
${E=-\cv{\tKW}\spr\frac{D\geod}{d\afp}}$,
${J=\cv{\ph}}\spr\frac{D\geod}{d\afp}$, and
${Q=\frac12\,\tens{Q}\mspace{0.8mu}(\frac{D\geod}{d\afp},\frac{D\geod}{d\afp})}$,
respectively, for any null geodesic. Namely, we have
\begin{equation}
\begin{gathered}
 E=\frac{\FKW\dot{\tKW}}{\accl^2(\xKW+\yKW)^2} \comma
 J=\frac{\GKW\dot{\ph}}{\accl^2(\xKW+\yKW)^2} \commae \\
 Q=\frac{\FKW\dot{\tKW}^2-\FKW^{-1}\dot{\yKW}^2
 +\GKW^{-1}\dot{\xKW}^2+\GKW\dot{\ph}^2}{2\accl^4(\xKW+\yKW)^4} \period
\end{gathered}
\end{equation}
In addition, $\frac{D\geod}{d\afp}$ has a null norm
\begin{equation}\label{normKW}
 -\FKW\dot{\tKW}^2+\FKW^{-1}\dot{\yKW}^2+\GKW^{-1}\dot{\xKW}^2+\GKW\dot{\ph}^2=0 \period
\end{equation}
The above four equations imply
\begin{equation}\label{tvecKW}
\begin{aligned}
 \dot{\tKW} &= E\accl^2(\xKW+\yKW)^2\FKW^{-1} \commae&
 \dot{\yKW} &= \epsilon_\yKW \accl^2(\xKW+\yKW)^2\sqrt{E^2-Q\FKW} \commae\\
 \dot{\ph}  &= J\accl^2(\xKW+\yKW)^2\GKW^{-1} \commae&
 \dot{\xKW} &= \epsilon_\xKW \accl^2(\xKW+\yKW)^2\sqrt{Q\GKW-J^2} \commae
\end{aligned}
\end{equation}
where ${\epsilon_\yKW,\epsilon_\xKW=\pm1}$ are the signs of $\dot{\yKW}$ and $\dot{\xKW}$, respectively.
The components of the tangent vector $\frac{D\geod}{d\afp}$
in the Robinson-Trautman coordinates (\ref{RTKW}) are thus given by
\begin{equation}\label{tvecRT}
\begin{aligned}
 \dot{\rRT} & = -\epsilon_\xKW\accl\sqrt{Q\GKW-J^2}-\epsilon_\yKW\accl\sqrt{E^2-Q\FKW} \commae\\
 \dot{\uRT} & = \frac{1}{\accl\rRT^2\FKW}\Bigl(\epsilon_\yKW\sqrt{E^2-Q\FKW}+E\Bigr) \commae\\
 \dot{\zRT} & = \frac{1}{\sqrt{2}\rRT^2}\Bigl(\epsilon_\yKW\sqrt{E^2-Q\FKW}\FKW^{-1}+E\FKW^{-1}\\
            &\mspace{60mu}-\epsilon_\xKW\sqrt{Q\GKW-J^2}\GKW^{-1}-iJ\GKW^{-1}\Bigr) \period
\end{aligned}
\end{equation}

We immediately observe that there exists a family of simple null geodesics
$\dot{\rRT}=\accl\,E$, ${\dot\zeta=0=\dot{u}}$, corresponding to the special
choice of the constants ${J=0=Q}$, ${\epsilon_\yKW=-\sign E}$.
These lie in the planes $\xKW,\ph=\text{constant}$, cf.\ \eqref{tvecKW},
and they are considered in Appendix~\ref{apx:AlgSpecDir}.

Now, we notice that $\dot\rRT$ remains finite at infinity $\scri$,
\begin{equation}
 \dot{\rRT} \lteq
  -\epsilon_\xKW\accl\sqrt{Q\GKW_\onscri-J^2}-\epsilon_\yKW\accl\sqrt{E^2-Q\FKW_\onscri}
  \equiv\rafpc\period
\end{equation}
This means that $\rRT$ and $\afp$ are asymptotically proportional,
\begin{equation}\label{rafpcdef}
  \rRT\lteq\rafpc\,\afp\period
\end{equation}
We can thus easily obtain the \emph{asymptotic behavior} of the above tangent vector
by expanding expressions \eqref{tvecRT} in powers of $1/\afp$
(assuming ${\rafpc\neq 0}$ because the geodesic approaches infinity).
We get%
\footnote{Let us note here that the vector ${\rRT^2\,\cvil{\rRT}=-\cvil{\Omega}}$
  is \vague{of the same order} as other coordinate vectors in the sense of expansion
  in~${1/\rRT}$, i.e., ${\rRT^2\,\cvil{\rRT} \sim \cvil{\zRT}\sim\cvil{\uRT}}$.
  More precisely, the vectors ${\cvil{\Omega},\,\cvil{\uRT},\,\cvil{\zRT}}$ are
  regular at $\scri$ in the sense of the tangent space of the conformal manifold with
  metric~\eqref{confmetric}, cf.\ Ref.~\cite{PenroseRindler:book2}.}
\begin{equation}\label{tangent}
  \frac{D\geod}{d\afp} \lteq \frac1{\rafpc\,\afp^2}\,\Bigl(\rRT^2\cv{\rRT}-c\,\cv{\zRT}
            -\bar c\,\cv{\bRT}-d\,\cv{\uRT}\Bigr)\commae
\end{equation}
with the constants ${c\in\complexn}$ and ${d\in\realn}$ related to the conserved quantities $E,\,J,\,Q$ by
\begin{equation}\label{cdEJQrel}
\begin{aligned}
 c & =  -\frac{1}{\rafpc\sqrt{2}}\Bigl(\epsilon_\yKW\sqrt{E^2-Q\FKW_\onscri}\FKW_\onscri^{-1}+E\FKW_\onscri^{-1} \\
   &\mspace{65mu} -\epsilon_\xKW\sqrt{Q\GKW_\onscri-J^2}\GKW_\onscri^{-1}-i J\GKW_\onscri ^{-1}\Bigr) \commae \\
 d & =  -\frac{1}{\rafpc \accl\FKW_\onscri}\Bigl(\epsilon_\yKW\sqrt{E^2-Q\FKW_\onscri}+E\Bigr) \period
\end{aligned}
\end{equation}
These constants are not independent. In fact, they satisfy the normalization condition \eqref{normKW}
which in Robinson-Trautman coordinates asymptotically reads
(recall Eqs.~(\ref{ae:PandH}) and (\ref{Hasympto}))
\begin{equation}\label{normRT}
  2\PRT_\onscri^{-2}c\bar c+2d+\frac{\Lambda}{3}\,d^2=0\period
\end{equation}

The expansion \eqref{tangent} of the tangent vector corresponds to the
asymptotic form of null geodesics ${\geod(\afp)}$ near~$\scri$
given by Eq.~\eqref{rafpcdef} and
\begin{equation}\label{asymp}
  \zRT =\zRT_\onscri+\frac{c}{\rafpc\,\afp} +\dots \comma
  \uRT =\uRT_\onscri+\frac{d}{\rafpc\,\afp} +\cdots \period
\end{equation}
The constants $\zRT_\onscri$, $\uRT_\onscri$ specify the \emph{position} at $\scri$
which particular geodesic \eqref{asymp} is approaching (or from which it is receding),
whereas $c$, $d$ represent the (spacetime) \emph{direction} along which $N_\onscri$ is reached.
The constant $\rafpc$ fixes the affine parameter $\afp$.
From Eq.~\eqref{rafpcdef} we see that if ${\rafpc>0}$ then $\rRT$ is growing
and geodesics are approaching the infinity for ${\afp\to+\infty}$ --- we will denote these as
outgoing. On the other hand, when ${\rafpc<0}$ then the geodesics approach $\scri$
(${\rRT\to\infty}$) as ${\afp\to-\infty}$, i.e., the coordinate
$\rRT$ is decreasing with a growing $\afp$.
The corresponding geodesics are ingoing: they \vague{start}
on $\scri$  and recede from this into finite regions of the spacetime.

Solving the normalization condition \eqref{normRT} we obtain
\begin{equation}\label{defd}
d+\frac{3}{\Lambda}= \EPS \frac{3}{\Lambda}\,\sqrt{1-\frac{\Lambda}{3}\,
  \frac{2c\bar c}{\PRT_\onscri^2}}\commae
\end{equation}
where ${\EPS=\pm 1}$. For any given  $c$
there are thus two real values of $d$, according to the sign of $\EPS$.
In fact, the above parameter $\EPS$ identifies whether the geodesic
is future or past oriented. To see this explicitly, let us consider the future-oriented
timelike vector $\cv{\uRT}$ near infinity
(${\cv{\uRT}\spr\cv{\uRT}=-\HRT\lteq \frac{\Lambda}{3}\,\rRT^2<0}$).
The projection of tangent vector \eqref{tangent} onto
$\cv{\uRT}$ is, using \eqref{defd},
\begin{equation}\label{proj}
  \frac{D\geod}{d\afp}\spr\cv{\uRT} \lteq -\rafpc\,\Big(1+\frac{\Lambda}{3}d\Big)
    = -\EPS\,\rafpc \,\sqrt{1-\frac{\Lambda}{3}\,
  \frac{2c\bar c}{\PRT_\onscri^2}}\period
\end{equation}
The geodesic is thus future or past oriented when
${\EPS\,\rafpc>0}$ or ${\EPS\,\rafpc<0}$, respectively.
Of course, it is physically natural to \emph{restrict ourselves to future-oriented}
geodesics only.
Without loss of generality we thus assume the identification
\begin{equation} \label{ident}
\sign\rafpc=\EPS \period
\end{equation}
Consequently, geodesics with ${\EPS =+1}$  are \emph{outgoing}
(reaching $\scri$ for ${\afp\to+\infty}$)
whereas those with ${\EPS =-1}$ are \emph{ingoing}
(starting at $\scri$ for ${\afp\to-\infty}$).

In order to find the radiative behavior of fields  near  $\scri$ we have to set
up an interpretation tetrad transported parallelly along a general asymptotic null
geodesic, and project the Weyl tensor  and
the tensor of electromagnetic field  onto this tetrad.   In fact,
in the following we will employ several orthonormal and null tetrads which will be
distinguished by specific labels in subscript.
We denote the vectors of a generic orthonormal tetrad  as
${\tG,\,\qG,\,\rG,\,\sG}$, where $\tG$ is a unit timelike
vector and the  remaining three are spacelike.
With this normalized tetrad we associate a null tetrad
${\kG,\,\lG,\,\mG,\,\bG}$,
\begin{equation}\label{NormNullTetr}
\begin{aligned}
  \kG &= \textstyle{\frac1{\sqrt{2}}} (\tG+\qG)\comma&
  \lG &= \textstyle{\frac1{\sqrt{2}}} (\tG-\qG)\commae\\
  \mG &= \textstyle{\frac1{\sqrt{2}}} (\rG-i\,\sG)\comma&
  \bG &= \textstyle{\frac1{\sqrt{2}}} (\rG+i\,\sG)\comma
\end{aligned}
\end{equation}
such that
\begin{equation}\label{NullTetrNorm}
   \kG\spr\lG = -1\comma
   \mG\spr\bG = 1\commae
\end{equation}
all other scalar products being zero.

The Weyl tensor is parametrized by five standard complex coefficients  $ \WTP{}{n}$, $n=0,1,2,3,4,$
defined as its specific components with respect to the above null tetrad, see Eq.~\eqref{PsiDef}.
Similarly, the tensor of electromagnetic field is parametrized by $\EMP{}{n}$, $n=0,1,2$,
see Eq.~\eqref{PhiDef}. The well-known transformation properties of coefficients $\WTP{}{n}$
and $\EMP{}{n}$ under null rotations, boost, and spatial rotation of the tetrad
are summarized in Appendix~\ref{apx:Transformations}.

To define a suitable \defterm{interpretation tetrad} ${\kI,\,\lI,\mI,\,\bI}$
we need to specify either its initial condition inside the spacetime, or its final
condition at timelike infinity $\scri$, in  a comparable way for all geodesics
approaching infinity along different directions.
We consider geodesics which reach the same point $N_\onscri$ at $\scri$, and thus
we prescribe the final condition there.
We naturally require that the null vector $\kI$ \emph{is proportional to the tangent vector \eqref{tangent}}
of the asymptotic null geodesic  \eqref{asymp},
\begin{equation}\label{kPandTangVect}
  \kI\lteq\frac{\EPS}{\rafpc}\,\frac{D\geod}{d\afp}\period
\end{equation}
We wish to \emph{compare} the radiation for all such null geodesics approaching
the given point at $\scri$, and it is thus
necessary to consider a unique and universal normalization of the affine
parameter $\afp$, and of the vector $\kI$. A natural and also the most
convenient choice is to \emph{keep the parameter $\rafpc$ fixed} ---  see an analogous
discussion in  \cite{KrtousPodolsky:2003a} near Eqs.~(5.6) and (5.9).
In fact, this  is equivalent to fixing the component ${\tens{p}\spr\norm}$
of the 4-momentum ${\tens{p}=\frac{D\geod}{d\afp}}$
at some large value of $\rRT$, i.e. at given the proximity of the
conformal infinity.

Following a general framework introduced in \cite{PenroseRindler:book2}, the null vector
$\lI$ of the interpretation tetrad now can be fixed asymptotically
by  normalization \eqref{NullTetrNorm} and the requirement that on $\scri$
\emph{the vector $\norm$ normal to the infinity belongs to $\kI\textdash\lI$ plane}.
Obviously, the direction of $\lI$ at a point $N_\onscri$ on $\scri$ thus uniquely depends on the choice of
the particular null geodesic \eqref{asymp} approaching infinity, i.e., on the specific vector $\kI$.
Remaining vectors ${\mI,\,\bI}$  cannot be prescribed canonically --- there is a freedom
in choice of their phase factor (a rotation in the transverse ${\mI\textdash\bI}$ plane).
Therefore, we have to find such physical quantities which
are invariant under this freedom. Obviously,
the \emph{moduli} $|\WTP{}{n}|$ and $|\EMP{}{n}|$ of the fields  at $\scri$
are independent of the specific choice of the vectors
${\mI,\,\bI}$.

To derive the field components in the above-defined interpretation tetrad
we start with the simple Robinson-Trautman null tetrad
${\kT,\,\lT,\,\mT,\,\bT}$ (see, e.g., \cite{Krameretal:book})
naturally adapted to the Robinson-Trautman coordinates \eqref{RTmetric}
\begin{equation}\label{RTtetrad}
\begin{aligned}
  \kT&=\;\cv{\rRT} \commae&&\\
  \lT&=-\textstyle{\frac{1}{2}}\HRT\,\cv{\rRT}+\cv{\uRT}\commae\\
  \mT&=\frac\PRT\rRT\,\cv{\bRT} \commae\\
  \bT&=\frac\PRT\rRT\,\cv{\zRT} \period
\end{aligned}
\end{equation}
Note that the vector $\kT$ is oriented along the double degenerate principal
null direction $\kG_1$, cf.\ Eq.~\eqref{PNDs}.
In this tetrad the only nontrivial components
$\WTP{\RoTr}{n}$ and  $\EMP{\RoTr}{n}$,
which represent the gravitational and electromagnetic fields, are
\begin{equation}\label{WeylRTTetr}
\begin{gathered}
  \WTP{\RoTr}{2}
  =-\Bigl(\mass + 2\,\charge^2\accl\, \xKW - \frac{\charge^2}\rRT\Bigr)\,\frac1{\rRT^3}
  \commae\\
  \WTP{\RoTr}{3} =
  -\frac3{\sqrt2}\,\frac{\accl\,\rRT}{\PRT}\,\WTP{\RoTr}{2}\comma
  \WTP{\RoTr}{4} =
  3\,\frac{\accl^2\,\rRT^2}{\PRT^2}\,\WTP{\RoTr}{2}\commae\\
  \EMP{\RoTr}{1} =  -\frac{\charge}{2\,\rRT^2}\comma
  \EMP{\RoTr}{2} =
  -\sqrt2\,\frac{\accl\,\rRT}{\PRT}\,\EMP{\RoTr}{1}\period
\end{gathered}
\end{equation}
The interpretation tetrad ${\kI,\,\lI,\mI,\,\bI}$ can be
obtained by performing two subsequent null rotations and a boost of this
Robinson-Trautman null tetrad \eqref{RTtetrad}. We first apply
\eqref{ae:lfixed}, then \eqref{ae:kfixed}, and finally \eqref{ae:boostrotation}
of  Appendix~\ref{apx:Transformations} with the parameters\footnote{%
For the particular case of geodesics with ${c=0}$ see Appendix~\ref{apx:AlgSpecDir}.}
\begin{equation}\label{para}
\begin{aligned}
  K&=-\Big(1+\frac{\Lambda}{6}d\Big)^{-1}\frac{ c}{\PRT\,\rRT}  \commae\\
  L&= \frac{\Lambda}{6}\frac{c\,\rRT}{\PRT}  \commae\\
  B&= \EPS\,\Big(1+\frac{\Lambda}{6}d\Big)  \comma\quad
  \Phi=0  \period
\end{aligned}
\end{equation}
The resulting null tetrad, using  relation \eqref{normRT},
then takes the following asymptotic form as ${\afp\to\EPS\infty}$,
\begin{equation}\label{genRTtetrad}
\begin{aligned}
  \kI&\lteq\EPS\frac1{\rafpc^2\afp^2}\,\left(\rRT^2\cv{\rRT}
     -c\,\cv{\zRT}-\bar c\,\cv{\bRT}-d\,\cv{\uRT}\right) \commae\\
  \lI&\lteq\EPS\frac{\Lambda}{6}\,\left(\rRT^2\cv{\rRT}
     +c\,\cv{\zRT}+\bar c\,\cv{\bRT}+\Big(d+\frac{6}{\Lambda}\Big)\,\cv{\uRT}\right) \commae\\
  \mI&\lteq\frac{\PRT_\onscri}{\rafpc\,\afp} \left(\,\frac{\Lambda}{6}\frac{c\,d}{\bar c}\,\cv{\zRT}
     +\Big(1+\frac{\Lambda}{6}d\Big)\,\cv{\bRT}-\frac{c}{\PRT_\onscri^2}\,\cv{\uRT}\right) \commae\\
  \bI&\lteq\frac{\PRT_\onscri}{\rafpc\,\afp} \left(\,\Big(1+\frac{\Lambda}{6}d\Big)\,\cv{\zRT}
     +\frac{\Lambda}{6}\frac{\bar c\,d}{c}\,\cv{\bRT}-\frac{\bar c}{\PRT_\onscri^2}\,\cv{\uRT}\right)
  \period
\end{aligned}
\end{equation}

The above vector $\kI$ is indeed obviously tangent to a general
asymptotic null geodesics \eqref{asymp}, and
satisfies the condition \eqref{kPandTangVect}. Moreover, the normal $\norm$
to  $\scri$, cf.\ Eq.~\eqref{gennorm}, belongs to the
plane spanned  by the two null vectors $\kI$ and $\lI$, as required,
\begin{equation}\label{Hasympt}
  \norm\lteq\frac{\EPS}{\sqrt2}\Bigl(\,
   \sqrt{-\frac{\Lambda}{6}}\;\rafpc\,\afp\;\kI
  -\sqrt{-\frac{6}{\Lambda}}\;\frac{1}{\rafpc\,\afp}\;\lI\,
  \Bigr)\period
\end{equation}
Notice that the projection of $\kI$ on $\norm$  is
\begin{equation}\label{kONn}
\kI\spr\norm\lteq \frac{\EPS}{\rafpc\,\afp}\sqrt{-\frac{3}{\Lambda}}\period
\end{equation}
For outgoing geodesics (${\EPS=+1}$, ${\afp\to+\infty}$) we indeed obtain ${\kI\spr\norm>0}$,
whereas for ingoing ones (${\EPS=-1}$, ${\afp\to-\infty}$) there is ${\kI\spr\norm<0}$.

Therefore, the  tetrad \eqref{genRTtetrad} is exactly the interpretation
tetrad suitable for analysis of the behavior of fields close to infinity~$\scri$.
As seen above, the Lorentz transformations from
the tetrad \eqref{RTtetrad} to \eqref{genRTtetrad} are given by two
subsequent null rotations and the boost with the parameters \eqref{para}.
Starting with the components \eqref{WeylRTTetr} in the
Robinson-Trautman frame we thus obtain, using
Eqs.~\eqref{ae:lfixedWeyl}, \eqref{ae:kfixedWeyl}, \eqref{ae:boostrotationWeyl} and
\eqref{ae:lfixedEM}, \eqref{ae:kfixedEM}, \eqref{ae:boostrotationEMF},
 the asymptotic form of the leading terms of
gravitational and electromagnetic fields
\begin{align}
\WTP{\intT}{4}&\lteq-\frac{3(\mass+2\charge^2\accl\,\xKW_\onscri)}{\rafpc\,\afp\,\PRT_\onscri^2}\,
\left[\,\accl\left(1+{\frac{\Lambda}{6}}\,d\right)-\frac{\Lambda\,\bar{c}}{3\sqrt2}\,\right]^2 \commae\notag\\
   \EMP{\intT}{2}&\lteq\ \frac{\EPS\,\charge}{\sqrt2\,\rafpc\,\afp\,\PRT_\onscri}\,
\left[\,\accl\left(1+{\frac{\Lambda}{6}}\,d\right)-\frac{\Lambda\,\bar{c}}{3\sqrt2}\,\right] \commae\label{FieldsOnScri}
\end{align}
where $\xKW_\onscri$ is the coordinate of the point $N_\onscri$ related to the coordinates
$\zRT_\onscri,\uRT_\onscri$ by
Eq.~\eqref{ae:relforx}. The other terms decrease faster in accordance with the well-known
\defterm{peeling behavior},  which is a consequence of the boost contained in \eqref{Hasympt}
that is infinite on $\scri$, cf.\ \cite{PenroseRindler:book2}.
Notice also that  the square of $\EMP{\intT}{2}$ gives the modulus of the Poynting
vector, $4\pi\abs{\EMS_\intT}\lteq\abs{\EMP{\intT}{2}}^2$,
defined in the interpretation tetrad \eqref{genRTtetrad}.
Interestingly, the dependence of ${\abs{\WTP{\intT}{4}}}$ and ${\abs{\EMP{\intT}{2}}^2}$
on the direction along which the point $N_\onscri$ at  infinity $\scri$ is
approached is \emph{exactly the same}.

Expressions \eqref{FieldsOnScri} are (formally) identical to Eqs.~(6.16) of \cite{KrtousPodolsky:2003a} in which radiation in the
$C$-metric spacetime with ${\Lambda>0}$ was investigated. Interestingly enough, one can also directly set
${\Lambda=0}$. The directional dependence given by the parameters $c$ and $d$ vanishes in this limit
(since $\lI$ becomes independent of $\kI$) and formulas \eqref{FieldsOnScri} can be compared with
the results  obtained  in classic work \cite{KinnersleyWalker:1970}  (see also \cite{FarhooshZimmerman:1979})
for accelerated black holes in asymptotically flat spacetime.

Nevertheless, the physical and geometrical meanings of Eqs.~\eqref{FieldsOnScri} is very different now since new
interesting and specific features occur for the ${\Lambda<0}$ case. The following sections will be devoted
to deeper description and analysis of the above result.

\section{Parametrizations of the null direction at $\scri$}
\label{sc:param}

For a physical understanding of expressions \eqref{FieldsOnScri},
as well as for explicit demonstration of fundamental differences
between radiation generated by accelerated black holes in spacetimes
with  ${\Lambda>0}$ and ${\Lambda<0}$, we introduce more convenient
parametrizations of the direction $\kI$ along which the infinity $\scri$
--- now timelike --- is approached.
To parametrize this radiation direction we first choose
a suitable \emph{reference tetrad} ${\tO,\,\qO,\,\rO,\,\sO}$
on $\scri$ which is orthonormal, adapted to the infinity,
\begin{equation}\label{qisn}
\qO=\norm\commae
\end{equation}
and with $\tO$ future oriented.
Otherwise the tetrad can be chosen arbitrarily.
A natural choice is to consider a tetrad closely related to the
Robinson-Trautman tetrad \eqref{RTtetrad}, namely,
\begin{equation}\label{RTreference}
 \kO=\sqrt{\frac{\HRT}{2}}\;\kT \commae\quad
 \lO=\sqrt{\frac{2}{\HRT}}\;\lT  \commae\quad
 \mO=\mT  \commae
\end{equation}
so that (cf.\ Eq.~\eqref{NormNullTetr})
\begin{equation}\label{RTtransfo}
\begin{aligned}
 \cv{\uRT}&=\sqrt{\HRT}\;\tO \commae\\
 \cv{\rRT}&=\frac{1}{\sqrt{\HRT}}\,\left(\,\tO+\qO\,\right)  \commae\\
  \cv{\zRT}&=\frac{1}{\sqrt2}\,\frac\rRT\PRT\,
   \left(\,\rO+i\,\sO \,\right)
 \period
\end{aligned}
\end{equation}

The null direction $\kI$ can be obtained (up to normalization)
from the above reference vector $\kO$ by a null
rotation \eqref{ae:lfixed}, and thus it can be parametrized
by a complex parameter $\R$ as%
\footnote{The special case ${\kI\propto\lO}$ formally corresponds to ${\R=\infty}$.
  Here and in the following the symbol $\propto$ means a proportionality with
  a \emph{positive} factor. Thanks to this convention we do not lose information
  about the orientation of related vectors.}
\begin{equation}\label{Rparam}
\kI\propto \kO + \bar \R\, \mO + \R\, \bO + \R \bar\R\, \lO\period
\end{equation}
Comparing this expression with Eq.~\eqref{genRTtetrad} we can relate $\R$ to the parameters $c$ and $d$,
\begin{equation}\label{cdRrel}
\begin{aligned}
c&=-\sqrt{-\frac6\Lambda}\,\PRT_\onscri\,\frac{\R}{1-\abs{\R}^2}\commae\\
d&=\frac6\Lambda\,\frac{\abs{\R}^2}{1-\abs{\R}^2}\period
\end{aligned}
\end{equation}
Now we may rewrite the directional pattern \eqref{FieldsOnScri} in terms of the parameter $\R$.
First, recall that there is no canonical way how to choose the phase of the
transverse null vectors ${\mI,\,\bI}$. Therefore, invariant information
independent of a choice of the interpretation tetrad
is contained only in the \emph{moduli} of fields components.
Substituting relations \eqref{cdRrel} into expressions~\eqref{FieldsOnScri} we obtain
\begin{equation}\label{resultR}
\begin{aligned}
  \abs{\WTP{\intT}{4}}&\lteq \frac{3\accl^2(\mass+2\charge^2\accl\,\xKW_\onscri)}{\rafpc\,\afp\,\PRT_\onscri^2}
    \, \frac{\abs{1-\R_1\bar\R}^2\abs{1-\R_2\bar\R}^2}{\bigl(1-\abs{\R}^2\bigr)^2}\commae\\
  \abs{\EMP{\intT}{2}}&\lteq \frac{\abs{\charge}\accl}{\sqrt2\,\rafpc\,\afp\,\PRT_\onscri}
    \, \frac{\abs{1-\R_1\bar\R}\abs{1-\R_2\bar\R}}{\bigabs{1-\abs{\R}^2}}\commae
\end{aligned}
\end{equation}
where
\begin{equation}\label{RofPNDs}
\R_1=0\comma\R_2=\sqrt{-\frac{\Lambda}{3}}\,\frac{\,\PRT_\onscri}{\accl}\period
\end{equation}

We have introduced here not only $\R_2$ but also a \vague{superfluous}
parameter ${\R_1=0}$. This is motivated by a general result
\cite{KrtousPodolsky:GEFADSI} concerning an asymptotic
structure of the fields when ${\Lambda<0}$.
In fact, the real parameters $\R_1$, $\R_2$ have an important physical meaning ---
they represent double-degenerate \defterm{principal null directions} (PNDs)
$\kG_1$ and $\kG_2$ (cf.\ Eq.~\eqref{PNDsKW}) at the infinity $\scri$.
Indeed, the specific complex parameter $\R$ representing a PND
with respect to the reference tetrad
has to satisfy quartic equation ${\WTP{}{0}=0}$ (see, e.g., \cite{Krameretal:book}),
i.e., using Eq.~\eqref{ae:lfixedWeyl},
\begin{equation}\label{PNDcond}
\R^4 \WTP{\refT}{4} + 4 \R^3 \WTP{\refT}{3} +
    6 \R^2 \WTP{\refT}{2} + 4 \R\, \WTP{\refT}{1} + \WTP{\refT}{0}=0\period
\end{equation}
This, in view of \eqref{RTreference},
\eqref{ae:boostrotationWeyl}, and \eqref{WeylRTTetr}, asymptotically reduces exactly to
\begin{equation}
(\R-\R_1)^2(\R-\R_2)^2=0\period
\end{equation}
Therefore, the first double-degenerate PND $\kG_1$ is indeed given by ${\R=\R_1}$,
whereas the second one, $\kG_2$, is given by ${\R=\R_2}$.

\begin{figure}
\includegraphics{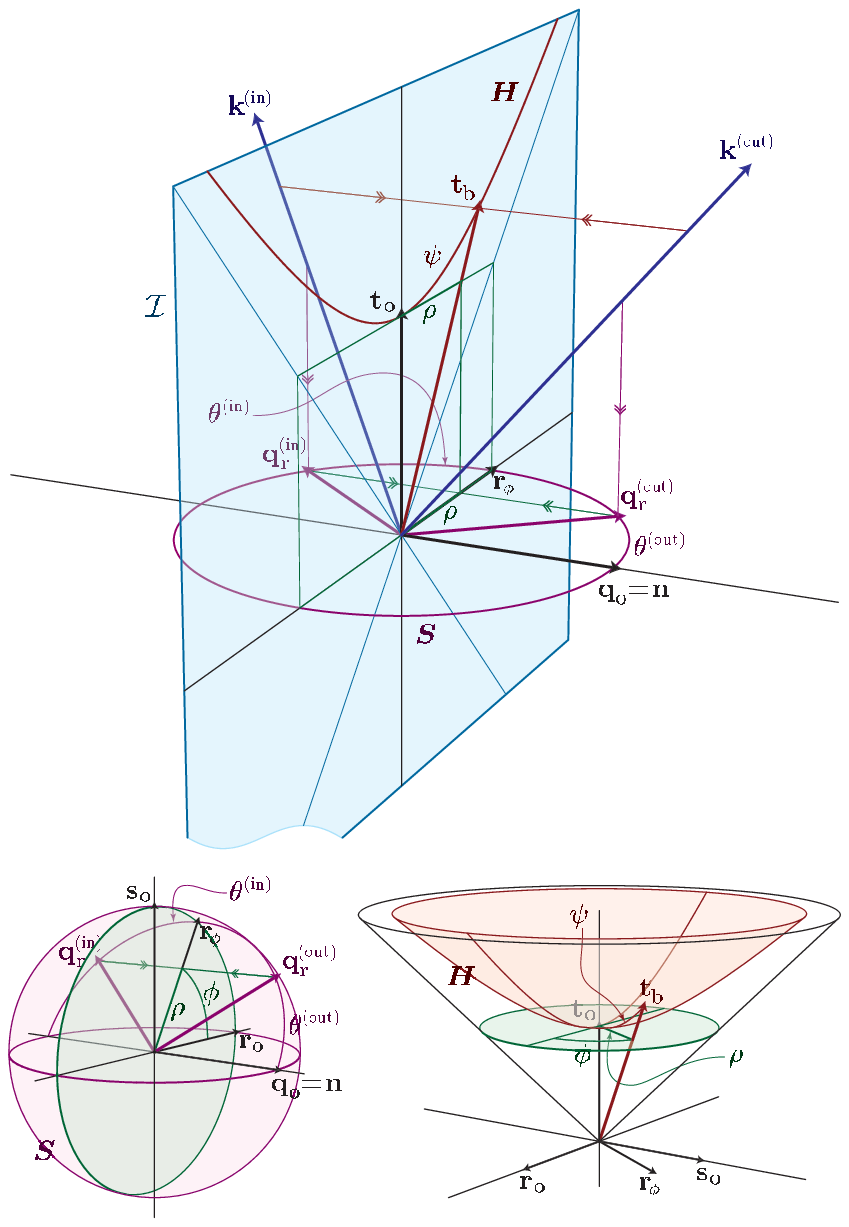}
\caption{\label{fig:nulldir}
Parametrizations of a null direction at timelike infinity $\scri$.
All null directions at a point $N_\onscri\in\scri$ can be characterized by
a future oriented null vector $\kG$.
According to their orientation ${\EPS=\sign(\kG\spr\norm)}$ with
respect to $\scri$, these can be divided into
\defterm{outgoing} and \defterm{ingoing} families --- see, e.g.,
vectors $\kG^{\mathrm{(out)}}$ and $\kG^{\mathrm{(in)}}$ in the figure.
(Special null directions \defterm{tangent} to $\scri$ will not be discussed here.)
The null direction is parametrized by a boost $\PSI$ and
an angle $\PHI$, or alternatively
by spherical angles ${\THT,\,\PHI}$.
These parametrizations are defined with respect to the
reference tetrad ${\tO,\,\qO,\,\rO,\,\sO}$. In the top diagram the vectors
${\tO,\,\qO,\,\rG_\PHI}$ (where ${\rG_\PHI=\cos\PHI\,\rO+\sin\PHI\,\sO}$)
are depicted, the remaining spatial direction $\sG_\PHI$ is suppressed.
In the bottom left diagram the timelike direction $\tO$ is suppressed
and all spatial directions are drawn. Finally, in the bottom right diagram
the spatial direction ${\qO=\norm}$ normal to $\scri$ is omitted.
The parameters ${\PSI,\,\PHI}$ specify the normalized orthogonal
projection $\tB$ (Eq.~\eqref{genproj}) of the null vector $\kG$ onto
$\scri$ by Eq.~\eqref{PSIPHIparam}.
All possible $\tB$ form a two-dimensional hyperboloid $\boldsymbol{H}$ drawn
in the bottom right diagram. This hyperboloid can be radially projected
onto a two-dimensional disk tangent to the vertex of the hyperboloid given by $\tO$. The disk can
be parametrized by radial coordinate ${\rho=\tanh\PSI}$, and angle $\PHI$.
Alternatively, the null direction can be characterized by
the normalized \emph{spatial} projection $\qR$ (Eq.~\eqref{genprojq})
of the null vector $\kG$ into the 3-space orthogonal to $\tO$.
The projection $\qR$ can be parametrized by spherical angles
${\THT,\,\PHI}$ with respect to the reference tetrad, see Eq.~\eqref{THTPHIparam}.
All spatial projections $\qR$ form a two-dimensional sphere $\boldsymbol{S}$
shown in the bottom left. This sphere can be
orthogonally projected onto $\scri$, where it again forms a two-dimensional disk
parametrized by ${\rho=\sin\THT}$, and~$\PHI$, cf.~Eq.~\eqref{THTPSIRrel}.
}%
\end{figure}

Instead of using the complex parameter $\R$ for
identification of the null direction $\kI$
we can introduce two real parameters with an obvious geometrical meaning.
First, we perform a normalized projection
of the null vector $\kI$ onto $\scri$,
defining thus the unit timelike vector $\tB$ tangent to the infinity:
\begin{equation}\label{genproj}
\tB=\frac{\kI-(\kI\spr\norm)\,\norm}{\abs{\kI\spr\norm}}\period
\end{equation}
Then $\tB$ represents the radiation direction along $\scri$ corresponding to
the null vector ${\kI\propto \,\tB+\EPS\,\norm}$.
We can characterize $\tB$ (and thus $\kI$) with respect to the reference tetrad as
\begin{equation}\label{PSIPHIparam}
  \tB = \cosh\PSI\;\tO + \sinh\PSI\,(\cos\PHI\;\rO + \sin\PHI\;\sO) \period
\end{equation}
The parameters ${\PSI,\PHI}$ are \defterm{pseudo-spherical} coordinates,
${\PSI\in[0,\infty)}$ corresponding to a \emph{boost},
and ${\PHI\in[0,2\pi)}$ being an \emph{angle}.
Their geometric meaning is visualized in Fig.~\ref{fig:nulldir}.
However, these parameters do not specify the null direction $\kI$ uniquely ---
there always exists \emph{one ingoing} and \emph{one outgoing} null direction with the same
parameters $\PSI$ and $\PHI$, which are distinguished by $\EPS$.

Substituting Eq.~\eqref{Rparam} into Eq.~\eqref{genproj}
and comparing with Eq.~\eqref{PSIPHIparam} we can express $\PSI$ and $\PHI$ in terms of $\R$ as
\begin{equation}\label{PSIPHIRrel}
  \tanh\PSI=\frac{2\abs{\R}}{1+\abs{\R}^2}\comma\PHI=-\arg\R\period
\end{equation}
Observing that the sign of the expression ${1-\abs{\R}^2}\propto{\kI\spr\norm}$ determines
whether $\kI$ is ingoing or outgoing, i.e.,
\begin{equation}\label{REPSrel}
  \EPS=\sign(1-\abs{\R}^2)\commae
\end{equation}
we can write down the inverse relations,
\begin{equation}\label{RPSIPHIrel}
  \R =
\begin{cases}
{\displaystyle \tanh\frac\PSI2\;\exp(-i\PHI)}&   \text{for $\kI$ outgoing (${\EPS\!=\!+1}$)}\commae
\vspace*{6pt}\\
{\displaystyle \coth\frac\PSI2\;\exp(-i\PHI)}&   \text{for $\kI$ ingoing (${\EPS\!=\!-1}$)}\commae
\end{cases}
\end{equation}
and also the relations to the parameters $c$ and $d$,
\begin{equation}\label{cdPSIPHIrel}
  -\EPS\sqrt{-\frac{\Lambda}{6}}\,\frac{2\bar c}{\PRT_\onscri}=\sinh\PSI\exp(i\PHI) \comma
  \EPS\Big(1+\frac{\Lambda}{3}\,d\Big)=\cosh\PSI \period
\end{equation}

\begin{figure*}
\includegraphics{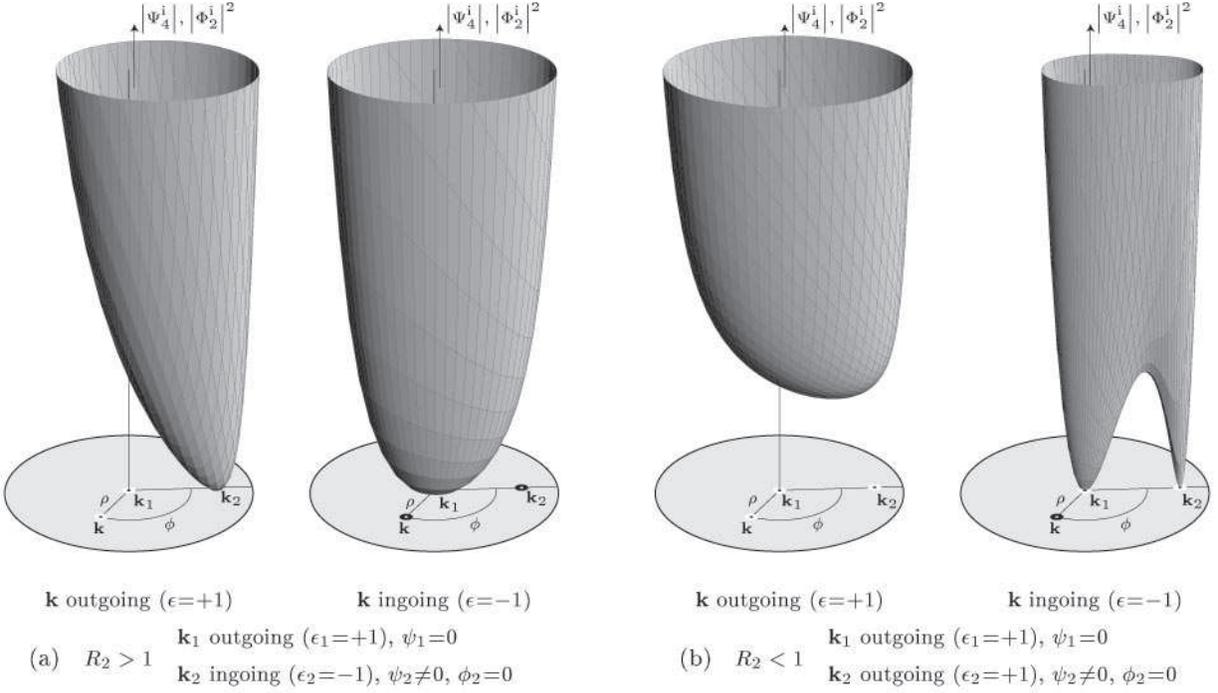}
\caption{\label{fig:dpr}
Possible directional patterns of radiation \eqref{resultPSIPHI}
[or, equivalently, \eqref{resultR}, \eqref{resultTHTPHI}]
which express the magnitude of the leading terms of gravitational or electromagnetic fields
as a function of a direction from which the point $N_\onscri$ at infinity $\scri$
is approached. The corresponding outgoing (${\EPS=+1}$, small white spot) or
ingoing (${\EPS=-1}$, black spot) null geodesic with a tangent vector $\kG$ are
parametrized by ${\rho=\tanh\PSI}$ and ${\PHI}$, cf.\ Fig.~\ref{fig:nulldir}. The patterns (a) apply to
points $N_\onscri$ in which the double degenerate PND $\kG_1$
(white spot) is oriented outwards from the universe whereas $\kG_2$
(black spot) inside it. The patterns (b) apply to points in which both the
PNDs are outgoing. The pattern (b) would also apply for points where both PNDs are ingoing,
only with exchanged words ``ingoing'' and ``outgoing''. The radiation completely vanishes along
directions which are exactly mirrored of the PNDs, with respect to $\scri$.}%
\end{figure*}

Substituting from \eqref{cdPSIPHIrel} into \eqref{FieldsOnScri} we obtain
\begin{equation}\label{resultPSIPHI}
\begin{aligned}
  \abs{\WTP{\intT}{4}}&\lteq\frac{\abs{\Lambda}}{4}\frac{(\mass+2\charge^2\accl\,\xKW_\onscri)}{\rafpc\,\afp}\\
    &\quad\times\, \abs{\R_2^{-1}\,(1+\EPS\cosh\PSI)-\EPS\sinh\PSI\,\exp(i\PHI)}^2 \commae\\
  \abs{\EMP{\intT}{2}}&\lteq\sqrt{\frac{\abs{\Lambda}}{24}}\, \frac{|\charge|}{\rafpc\,\afp}\\
    &\quad\times\, \abs{\R_2^{-1}\,(1+\EPS\cosh\PSI)-\EPS\sinh\PSI\,\exp(i\PHI)}
    \period
\end{aligned}
\end{equation}
The dependence of the fields on the parameters $\PSI$ and $\PHI$ is shown
in Fig.~\ref{fig:dpr}.

There exists yet another natural possibility how to
characterize the null direction at the infinity.
Instead of decomposing the propagation vector $\kI$
into the component normal to $\scri$ and the transverse timelike vector
$\tB$ tangent to $\scri$, we may alternatively consider its normalized \emph{spatial}
projection $\qR$, where by the spatial projection we mean a projection
to a suitable three-dimensional space,
say that orthogonal to $\tO$ (see Fig.~\ref{fig:nulldir}).
This spatial propagation vector
\begin{equation}\label{genprojq}
\qR=\frac{\kI+(\kI\spr\tO)\,\tO}{|\kI\spr\tO|}\commae
\end{equation}
such that ${\kI\propto\tO+\qR}$,
is naturally characterized by \emph{spherical} angles
$\THT$ and $\PHI$ with respect to the spatial vectors ${\qO,\,\rO,\,\sO}$ of
the reference tetrad, namely
\begin{equation}\label{THTPHIparam}
  \qR = \cos\THT\;\qO + \sin\THT\,(\cos\PHI\;\rO + \sin\PHI\;\sO) \period
\end{equation}
Obviously, this is more convenient for a unified description of
both outgoing (${\EPS=+1}$) and ingoing (${\EPS=-1}$) null geodesics.
The former are parametrized by ${\THT\in[0,\frac{\pi}{2})}$, the latter by
${\THT\in(\frac{\pi}{2},\pi]}$. Comparing with the previous parametrizations
of the null direction $\kI$ we obtain
\begin{equation}\label{THTPSIRrel}
\begin{aligned}
  \sin\THT&=\tanh\PSI=\frac{2\abs{\R}}{1+\abs{\R}^2}\equiv\rho\commae\\
  \cos\THT&=\EPS\,\sech\PSI=\frac{1-\abs{\R}^2}{1+\abs{\R}^2}\commae\\
  \tan\THT&=\EPS\,\sinh\PSI=\frac{2\abs{\R}}{1-\abs{\R}^2}\period
\end{aligned}
\end{equation}
The parameter $\rho$ is used in Figs.~\ref{fig:nulldir}, \ref{fig:dpr} and \ref{fig:dpr00}.
The inverse relation
\begin{equation}\label{RTHTPHIrel}
  \R = \tan\frac\THT2\;\exp(-i\PHI)\commae
\end{equation}
and analogous expression Eq.~\eqref{RPSIPHIrel}, show that $\R$
is actually a \defterm{stereographic} parametrization of $\qR$, and
\defterm{Lorentzian stereographic} parametrization of $\tB$.

Expressing the fields \eqref{resultPSIPHI} using $\THT$ and $\PHI$ we get
\begin{equation}\label{resultTHTPHI}
\begin{aligned}
  \abs{\WTP{\intT}{4}}&\lteq\frac{\abs{\Lambda}}{4}\frac{(\mass+2\charge^2\accl\,\xKW_\onscri)}{\rafpc\,\afp\,\cos^2\THT}\\
    &\quad\times\, \abs{\R_2^{-1}\,(1+\cos\THT)-\sin\THT\,\exp(i\PHI)}^2 \commae\\
  \abs{\EMP{\intT}{2}}&\lteq\sqrt{\frac{\abs{\Lambda}}{24}}\, \frac{|\charge|}{\rafpc\,\afp\,\abs{\cos\THT}}\\
    &\quad\times\, \abs{\R_2^{-1}\,(1+\cos\THT)-\sin\THT\,\exp(i\PHI)} \period
\end{aligned}
\end{equation}

We have thus presented the directional radiation pattern at the anti--de~Sitter infinity
using three suitable parametrizations of the null direction
along which the infinity is approached, namely, Eqs.~\eqref{resultR}, \eqref{resultPSIPHI}, and \eqref{resultTHTPHI}.
The pattern is depicted in  Fig.~\ref{fig:dpr}.
Now we can proceed with its physical interpretation.

\section{Analysis of the radiation pattern}
\label{sc:Analysis}

First, we observe that the radiation \vague{blows up} for directions with ${\abs{\R}=1}$,
i.e., ${\PSI\to\infty}$, ${\THT=\pi/2}$, ${\rho=1}$. These are null directions \emph{tangent} to
the infinity $\scri$, and thus they do not represent a direction
of any outgoing or ingoing geodesic approaching the infinity from
the \vague{interior} of the spacetime. The reason for this divergent behavior of the radiation
is purely kinematic: by imposing the \vague{comparable} final
conditions for the interpretation tetrad (cf.\ the discussion after Eq.~\eqref{kPandTangVect})
we have fixed the projection of the 4-momentum $\mom\propto\kI$ onto the normal $\norm$.
Clearly, this condition leads to an \vague{infinite} rescaling of $\kI$ if
$\kI$ is tangent to $\scri$, i.e., orthogonal to $\norm$.
Such rescaling results in the above divergence of $\abs{\WTP{\intT}{4}}$
and $\abs{\EMP{\intT}{2}}$.

This divergence at ${\abs{\R}=1}$ actually splits the radiation pattern
into two components ---  the radiation pattern for \emph{outgoing} geodesics,
${\abs{\R}<1}$, and to the pattern for \emph{ingoing} geodesics, ${\abs{\R}>1}$,
cf.\ \eqref{REPSrel}.
These two different patterns correspond to Eq.~\eqref{resultPSIPHI} with ${\EPS=+1}$
and ${\EPS=-1}$, respectively. They are depicted in Fig.~\ref{fig:dpr} as
separate diagrams.

From Eqs.~\eqref{resultR} it can immediately be observed that radiation completely vanishes,
${\abs{\WTP{\intT}{4}}=0=\abs{\EMP{\intT}{2}}}$, along specific null directions
with ${\R=\R_\m}$ satisfying
\begin{equation}\label{zerosR}
\R_\m=\frac1{\R_1}\quad\text{or}\quad\R_\m=\frac1{\R_2}\period
\end{equation}
In fact, the direction given by $1/\R_n$, ${n=1,2}$ is
the \defterm{mirrored reflection} of the PND $\kG_n$ with respect to $\scri$:
using Eqs.~\eqref{REPSrel}, \eqref{RPSIPHIrel} we find that both $1/\R_n$ and $\R_n$ correspond
to the same ${\PSI=\PSI_n}$ (and ${\PHI=0}$) but with the opposite  $\EPS$.
\emph{The radiation thus vanishes along mirrored reflections of the PNDs.}%
\footnote{In general, a mirrored reflection of the direction
$\R$ is ${1/{\bar\R}}$, but the PNDs $\R_1$ and $\R_2$ are real, see Eq.~\eqref{RofPNDs}.}

In terms of pseudo-spherical parameters $\PSI$, $\PHI$ we find,
cf.\ also \ Eqs.~\eqref{resultPSIPHI}, that the radiation vanishes along
outgoing (${\EPS=+1}$) null geodesics such that
\begin{equation}\label{zeros1}
\coth\frac{\PSI_\m}{2}=\R_2\comma\PHI_\m=0 \commae
\end{equation}
or along ingoing (${\EPS=-1}$) geodesics  given by
\begin{equation}\label{zeros2}
\begin{aligned}
\tanh\frac{\PSI_\m}{2}&=\R_1\;\Rightarrow\; \PSI_\m=0\comma\;\text{or}\\
\tanh\frac{\PSI_\m}{2}&=\R_2\comma\PHI_\m=0\period\qquad
\end{aligned}
\end{equation}
Clearly, only one of the conditions \eqref{zeros1}, \eqref{zeros2} involving $\R_2$
can be satisfied for a given value of $R_2$. Therefore, further description of the radiation
pattern necessarily depends on the \emph{specific algebraic structure
of the spacetime at a given point $N_\onscri$ at $\scri$},
in particular on the \emph{orientation} of  PNDs.

The PNDs have  explicit form, cf.\ Eqs.~\eqref{Rparam}, \eqref{RofPNDs},
\begin{equation}\label{PNDs}
\begin{aligned}
   \kG_1 &\propto {\frac{1}{\sqrt 2}}\,\left(\,\tO+\qO\,\right)=\kO\propto\kT   \commae\\
   \kG_2 &\propto {\frac{1}{\sqrt 2}}\,
     \left(\tO+\frac{1-\R_2^2}{1+\R_2^2}\,\qO
     +\frac{2\R_2}{1+\R_2^2}\,\rO\right) \period
\end{aligned}
\end{equation}
Using the relation \eqref{RPSIPHIrel} they can be parametrized as
\begin{equation}\label{psis}
\begin{aligned}
\PSI_1&=0    \commae\\
\tanh\frac{\PSI_2}{2}&=\left\{
\begin{matrix}
\>\Source  & \text{ for } \ \Source\le 1\cr
\noalign{\medskip}
\>1/\Source & \text{ for } \ \Source\ge 1\cr
\end{matrix}
\right.\commae\quad
\PHI_2=0 \period
\end{aligned}
\end{equation}
By inspecting Eqs.~\eqref{PNDs} we observe that the first PND $\kG_1$
always points along the normal ${\norm=\qO}$, i.e., \emph{outside the universe}.
However, for the second PND $\kG_2$ there are
distinct possibilities according to whether ${\R_2\lessgtr1}$.
At points on $\scri$ where ${\R_2<1}$ the vector $\kG_2$ is \emph{outgoing}
(${\EPS_2=+1}$), i.e., oriented \emph{outside} the universe.
In the regions where ${\R_2>1}$ it is \emph{ingoing}
(${\EPS_2=-1}$), oriented \emph{inside} the universe.
At special points where $\R_2=1$ the PND ${\kG_2\propto(\tO+\rO)}$
has no component along $\norm$; it is \emph{tangent} to $\scri$.

Which of these three alternatives can occur depends on values of the parameters
describing the spacetime. Before we continue with a discussion of the different possibilities,
let us note that the three possible regions of $\scri$ with the distinct structure of PNDs
\emph{exactly coincide} with regions of different \emph{characters
of the Killing vector field} $\cvil{\tKW}$. Recalling \eqref{infinity}, the value
of the metric function $\FKW$ at a given point $N_\onscri$ on $\scri\,$ is
${\FKW_\onscri=\FKW\vert_{\yKW=-\xKW_\onscri}}$.
Considering  Eqs.~\eqref{ae:KWFGQ}, \eqref{ae:PandH}, and  \eqref{RofPNDs} we obtain
\begin{equation}\label{FtoSource}
\FKW_\onscri=\frac{\R_2^2-1}{{\PRT^2_\onscri}}=(\R_2^2-1)\,\GKW_\onscri\commae
\end{equation}
which demonstrates the relation between the structure of PNDs  and the character
of the spacetime near infinity. If $\R_2<1$ then $\FKW_\onscri<0$ and the Killing vector
$\cvil{\tKW}$ near $\scri$ is spacelike. If, instead, $\R_2>1$ then $\FKW_\onscri>0$ and $\cvil{\tKW}$ is a timelike
 Killing vector field --- the region near $\scri$ is thus \emph{static}. The above two domains of infinity
are separated by the \emph{Killing horizon} consisting of points for which
$\R_2=1$, where the Killing vector is null.
Note that the Killing horizons
may indeed extend (for sufficiently large acceleration)
to the conformal infinity, which is a specific property of anti--de~Sitter
$C$-metric.

\subsection{A single accelerated black hole}
\label{subsc: small A}

\begin{figure}
\includegraphics{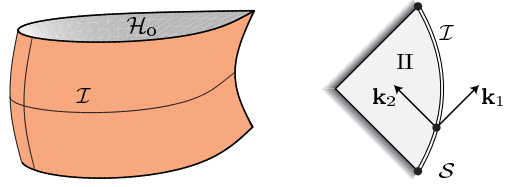}
\caption{\label{fig:CAdSIscri}
When ${\accl<\sqrt{-\Lambda/3}}$, the $C$-metric represents a single, uniformly
accelerated black hole. The region near infinity $\scri$ is
everywhere static. The first PND $\kG_1$ is oriented outside the
universe, whereas the second one $\kG_2$ points inside it.}%
\end{figure}

We now discuss the case when the acceleration parameter $\accl$ is
small, ${\accl<\sqrt{-\Lambda/3}}$. As explained in Sec.~\ref{ssc:single},
the $C$-metric then describes a single uniformly accelerated black hole in an anti--de~Sitter universe.
Its global structure for a constant $\ph$ is visualized  in Fig.~\ref{fig:CAdSI}.
There are \emph{no} Killing horizons extending to $\scri$ (which would correspond to
${\FKW_\onscri=0}$, i.e., ${\R_2=1}$)
since using \eqref{RofPNDs}, \eqref{ae:cosi}, and ${0\le\GKW\le1}$ there is
\begin{equation}\label{noKill}
\R_2>\PRT_\onscri=\frac{1}{\sqrt{\GKW_\onscri}}\ge1 \commae
\end{equation}
for all $\xKW\in[\xf,\xb]$. Accordingly, the region near infinity $\scri$
is \emph{everywhere static}. We thus find that the first PND $\kG_1$ is always oriented
\emph{outside}, whereas the second one  $\kG_2$ is always oriented \emph{inside}
the universe, see \eqref{PNDs} and Fig.~\ref{fig:CAdSIscri}.

The corresponding radiation pattern is shown in  Fig.~\ref{fig:dpr}(a).
There exists \emph{just one} direction
along which the \emph{outgoing} radiation (${\EPS=+1}$, ${\abs{\R}<1}$) vanishes,
namely the direction $\R_\m=1/R_2$.
It is the mirrored reflection of the ingoing PND $\kG_2$
(see the left part of Fig.~\ref{fig:dpr}(a)).
In terms of pseudo-spherical parameters
the direction is described by Eq.~\eqref{zeros1},
i.e., ${\PSI_\m=\PSI_2}$ and ${\PHI_\m=\PHI_2=0}$, where
$\PSI_2$ --- the boost parameter characterizing $\kG_2$ ---
is given by Eq.~\eqref{psis}.
The radiation pattern for \emph{ingoing} radiation
(${\EPS=-1}$, ${\abs{\R}>1}$) is visualized in the right part
of Fig.~\ref{fig:dpr}(a). Again, there exists just one direction of vanishing
radiation given by ${\R_\m=1/R_1=\infty}$ (i.e., ${\PSI_\m=\PSI_1=0}$, cf.\ Eq.~\eqref{zeros2}),
which is also the mirrored reflection of a PND,
this time of the outgoing $\kG_1$.

\subsection{A pair of accelerated black holes}
\label{subsc: big A}

\begin{figure}
\includegraphics{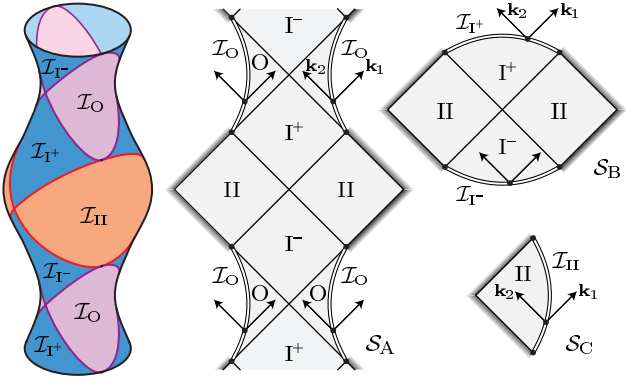}
\caption{\label{fig:CAdSIIscri}
For large values of ${\accl>\sqrt{-\Lambda/3}}$,
the $C$-metric represents pairs of accelerated
black holes. The conformal infinity $\scri$ shown on the left is
divided by horizons $\horc$ and $\hora$ into several distinct
domains: $\scriO$ and $\scriII$ are static, whereas $\scriIp$ and
$\scriIm$ are nonstatic. Moreover, in the regions $\scriO$
(corresponding to sections $\sect{A}$) and $\scriII$ (corresponding to
$\sect{C}$) the PND $\kG_1$ is oriented outwards, whereas the PND $\kG_2$
is oriented inwards. In $\scriIp$ (sections $\sect{B}$) both $\kG_1$ and $\kG_2$
point outside the universe, in $\scriIm$ both the PNDs point inside it.}%
\end{figure}

A more interesting but also more complicated situation occurs when
${\accl>\sqrt{-\Lambda/3}}$. In this case the $C$-metric
represents pairs of uniformly accelerated black holes in
an anti--de~Sitter universe, as indicated in
Figs.~\ref{fig:CAdSII}--\ref{fig:CAdSIIC}, see Sec.~\ref{ssc:pair}.
There are (outer) black-hole horizons $\horo$,
acceleration horizons $\hora$, and cosmological
horizons $\horc$, see Fig.~\ref{fig:CAdSII}(a).
At $\scri$, the horizons $\hora$ and $\horc$ can be identified by ${\R_2=1}$.
They separate various static and nonstatic regions of $\scri$, and
simultaneously the domains of infinity with different structure of the PNDs,
as shown in Fig.~\ref{fig:CAdSIIscri}.
The vector $\kG_1$ is always oriented outside the universe.
In the static domains of $\scri$ where ${\R_2>1}$,
denoted as $\scriO$ and $\scriII$,
$\kG_2$ is oriented inside it; see \eqref{PNDs}.
These domains of the infinity can be reached through
the sections $\sect{A}$ and $\sect{C}$, cf.\ Figs.~\ref{fig:CAdSIIA}, \ref{fig:CAdSIIC}.
On the other hand, in the domain where ${\R_2<1}$, denoted as $\scriIp$
(accessible through $\sect{B}$, Fig.~\ref{fig:CAdSIIB}),
both PNDs are oriented outside the spacetime.

In each of these  regions the radiation pattern \eqref{resultPSIPHI}
is thus different. In particular, it admits a different number of
directions along which the radiation vanishes.
Recalling that the radiation vanishes along mirrored reflections of PNDs,
we see that in the  static regions $\scriO$ and $\scriII$
there is just one \emph{outgoing} (${\EPS=+1}$)
direction along which the radiation vanishes,
as in the previous case of a single black hole.
This is the mirrored reflection of $\kG_2$, ${\R_\m=1/\R_2}$
(${\PSI_\m=\PSI_2}$, ${\PHI_\m=\PHI_2=0}$, cf.\ Eq.~\eqref{zeros1}).
The corresponding directional pattern
is again given by the left part of Fig.~\ref{fig:dpr}(a).
However, in the nonstatic region $\scriIp$ there is no \emph{outgoing} direction
along which the radiation vanishes because mirrored reflections of both PNDs are ingoing.
In other words, the condition~\eqref{zeros1} cannot be satisfied because ${\R_2<1}$.
The radiation pattern for outgoing directions for this case is shown in
the left part of Fig.~\ref{fig:dpr}(b).

Of course, the number of null directions with vanishing radiation in the pattern
for \emph{ingoing} geodesics (${\EPS=-1}$) is  complementary.
In the domains $\scriO$ and $\scriII$ with ${\R_2>1}$ there is again just one zero,
now given by the mirrored reflection of $\kG_1$ (${\R_\m=\infty}$, ${\PSI_\m=\PSI_1=0}$);
see the right part of Fig.~\ref{fig:dpr}(a).
On the other hand, in the domain $\scriIp$
there are exactly two zeros for ingoing radiation
given by mirrored reflections of both PNDs
(${\R_\m=\infty}$, ${\PSI_\m=0}$, and ${\R_\m=1/\R_2}$,
${\PSI_\m=\PSI_2}$, ${\PHI_\m=0}$, cf.\ Eq.~\eqref{zeros2})
as shown in the right part of Fig.~\ref{fig:dpr}(b).

In addition, there is also another domain with $\R_2<1$, namely
the domain $\scriIm$; see Fig.~\ref{fig:CAdSIIscri}.
Here both PNDs are oriented inside the spacetime.
However, the region $\Iminus$ of the spacetime and its infinity $\scriIm$ are
\emph{not covered} by the same map of
Robinson-Trautman coordinates as that used above.
We have to introduce another \vague{time-reversed}
map to cover the white domain in Fig.~\ref{fig:CAdSIIB}.
Still, the directions of vanishing radiation are given by
\emph{mirrored directions} of the PNDs at the infinity,
similarly to the cases discussed above.
The mirrored directions of the PNDs
are both outgoing so that the radiation pattern
for outgoing geodesics contains \emph{two} zeros
(cf.\ the right part of Fig.~\ref{fig:dpr}(b)),
whereas the pattern for ingoing geodesics does not have any zero directions,
cf.\ the left part of Fig.~\ref{fig:dpr}(b).

To summarize, the directional patterns of outgoing radiation
(Eqs.~\eqref{resultPSIPHI} with ${\EPS=+1}$, or Eqs.~\eqref{resultR} with ${\abs{\R}<1}$)
in the domains $\scriO$, $\scriIp$, $\scriIm$, and $\scriII$ of the conformal infinity
are given by the left~(a), left~(b), right~(b), and left~(a) parts of Fig.~\ref{fig:dpr}, respectively.
For ingoing radiation (Eqs.~\eqref{resultPSIPHI} with ${\EPS=-1}$, or Eqs.~\eqref{resultR} with ${\abs{\R}>1}$)
these are given by the right~(a),  right~(b), left~(b), and right~(a) parts of Fig.~\ref{fig:dpr}, respectively.

\subsection{Vanishing acceleration}
\label{ssc:vanishingA}

Finally, we briefly describe the character of  radiation
when the \emph{acceleration vanishes}, in which case the spacetime describes a single nonaccelerating
black hole in an anti--de~Sitter universe (\vague{Reissner-Nordstr\"om-anti--de~Sitter solution}).
It is thus spherically symmetric with surfaces ${\tKW,\yKW=\text{constant}}$ being the orbits of
the rotational group, and radial directions being contained in sections ${\xKW,\ph=\text{constant}}$.
From ${\accl=0}$ it follows ${\R_2=\infty}$,
cf.\ Eq.~\eqref{RofPNDs}, and  the above general expressions simplify.
In particular for ${\accl=0}$ not only
$\kT\propto\kG_1\propto{\frac{1}{\sqrt2}}(\tO+\qO)$, but also
$\lT\propto\kG_2\propto{\frac{1}{\sqrt2}}(\tO-\qO)$ is a PND,
see Eq.~\eqref{PNDs}; this is consistent with relations
\eqref{WeylRTTetr} in which only the components
$\WTP{\RoTr}{2}$ and $\EMP{\RoTr}{1}$ remain nonvanishing.
At the infinity $\scri$ the PNDs are mutually mirrored reflections, $\kG_1$ oriented outside the universe and
$\kG_2$ inside it, parametrized by ${\PSI_1=0=\PSI_2}$, see Eqs.~\eqref{PNDs}, \eqref{PSIPHIRrel}.
Both the PNDs point in radial directions of the spherical symmetry.
Moreover, since ${\R_2>1}$, it follows from relation \eqref{FtoSource} that the region near
$\scri$ is always static, without the Killing horizon extending there.

\begin{figure}
\includegraphics{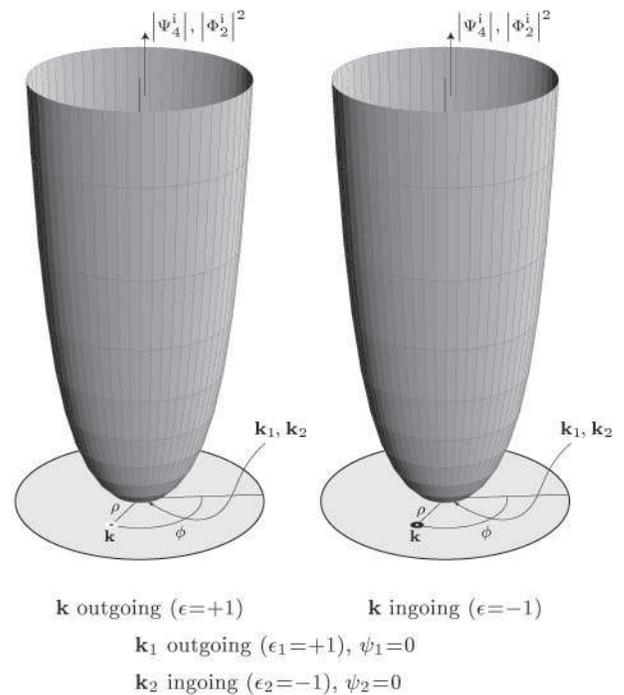}
\caption{\label{fig:dpr00}
The directional patterns of outgoing/ingoing radiation are the same and
axially symmetric when $A=0$. In such a case the PNDs $\kG_1$ and $\kG_2$ are
mutual mirror images under reflection with respect to $\scri$. Therefore,
there is exactly one direction given by $\PSI=0$ along which the radiation
vanishes, both for outgoing and ingoing null geodesics.}%
\end{figure}

For vanishing acceleration the expressions
\eqref{resultPSIPHI} and \eqref{resultTHTPHI} reduce to
\begin{equation}\label{DirCharAzero}
\begin{aligned}
   \abs{\WTP{\intT}{4}}&
     \lteq\frac{\abs{\Lambda}}{4}\frac{\mass}{\rafpc\,\afp}\,\sinh^2\PSI
     =    \frac{\abs{\Lambda}}{4}\frac{\mass}{\rafpc\,\afp}\,\tan^2\THT \commae\\
   \abs{\EMP{\intT}{2}}&
     \lteq\sqrt{\frac{\abs{\Lambda}}{24}}\,\frac{\abs{\charge}}{\rafpc\,\afp}\,\sinh\PSI
     =    \sqrt{\frac{\abs{\Lambda}}{24}}\,\frac{\abs{\charge}}{\rafpc\,\afp}\,\abs{\tan\THT}\period
\end{aligned}
\end{equation}
The corresponding directional pattern of radiation, shown in
Fig.~\ref{fig:dpr00}, is axially symmetric and independent of
$\EPS$, i.e., it is the same both for ingoing and outgoing null geodesics.
Interestingly, even for a nonaccelerated black hole there is thus radiation on $\scri$ along
\emph{any nonradial} null direction, i.e.,
except for ${\PSI=0}$, ${\EPS=\pm1}$, which corresponds to both PNDs.
This may seem quite surprising since the region near infinity is \emph{static}.
It is a completely new specific feature: for ${\Lambda>0}$ a
generic, nonradial observer near $\scri^+$ would
also detect radiation generated by nonaccelerated black holes
\cite{KrtousPodolsky:2003a}, but the region near infinity is nonstatic.
In asymptotically flat spacetimes (${\Lambda=0}$) there is no radiation on $\scri^+$
from black holes with ${\accl=0}$ \cite{KinnersleyWalker:1970}, which, remarkably, also follows from expression \eqref{FieldsOnScri}.

\section{Summary}

It was observed already in the 1960s by Penrose \cite{Penrose:1964,Penrose:1965,Penrose:1967}
that radiation  is defined \vague{less invariantly}  in spacetimes  with
a non-null $\scri$. Recently we have analyzed the $C$-metric solution with
${\Lambda>0}$ \cite{KrtousPodolsky:2003a} and found an explicit directional pattern of
radiation. This fully characterizes the radiation from uniformly accelerated black
holes near the de~Sitter-like conformal infinity $\scri^+$, which has a spacelike character.
Here we have completed the picture by investigating  the radiative properties
of the $C$-metric with a negative cosmological constant. This
exact solution of the Einstein-Maxwell equations with ${\Lambda<0}$ represents a
spacetime in which the radiation is generated either by a (possibly charged)
single black hole or pairs of black holes uniformly accelerated in
an anti--de~Sitter universe.

We have analyzed the asymptotic behavior of the gravitational
and electromagnetic fields near the conformal infinity $\scri$, which has a timelike character.
The leading components of the fields have been expressed in a suitable parallelly
transported interpretation tetrad. These components are inversely proportional
to the affine parameter of the corresponding null geodesic. In addition, an explicit
formula \eqref{resultPSIPHI} [or, equivalently, Eqs.~\eqref{resultR} and \eqref{resultTHTPHI}]
which describes the directional pattern of radiation has
been derived: it expresses the dependence of the field magnitudes on
spacetime directions from which a given point $N_\onscri$ at
infinity $\scri$ is approached. This specific directional
characteristic supplements the peeling property, completing thus
the asymptotic behavior of gravitational and electromagnetic
fields near infinity $\scri$ with a timelike character.

We have demonstrated that the situation is much more complicated
in the anti--de~Sitter case than in the case ${\Lambda>0}$.
The new specific feature is that timelike conformal
infinity $\scri$ is, in general,  divided by Killing horizons into several
static and nonstatic regions with a different structure of (double degenerate)
principal null directions. In these distinct domains of infinity the directional
patterns of radiation differ. For example, there are
different numbers of geometrically privileged directions (namely one, two, or none)
in which the radiation vanishes completely. These exactly correspond to
mirrored directions of principal null direction, with respect to
$\scri$. Accordingly, there exists an asymmetry between outgoing and ingoing
radiation patterns in all the domains.

As in the ${\Lambda>0}$ case \cite{KrtousPodolskyBicak:2003}, it seems plausible that a
general structure of the radiation pattern at conformal infinity
depends only on the PNDs there, i.e., it
is given by the algebraic (Petrov) type of the
spacetime. This hypothesis will be proven elsewhere \cite{KrtousPodolsky:GEFADSI}.

\begin{acknowledgments}
The work was supported in part by the grants GA\v{C}R 202/02/0735 and GAUK 166/2003
of the Czech Republic and Charles University in Prague.
The stay of M.O. at the Institute of Theoretical Physics in Prague
was enabled by financial support from Fondazione Angelo Della Riccia (Firenze).
\end{acknowledgments}

\appendix

\section{Radiation along some particular null geodesics}
\label{apx:AlgSpecDir}

Here we concentrate on a geometrically privileged family of special geodesics
which asymptotically take the form Eq.~\eqref{asymp}  with ${c=0}$. It follows from
\eqref{defd} that this corresponds either to outgoing null geodesics with ${d=0}$ or
ingoing ones with ${d=-6/\Lambda}$.

In fact, the geodesics ${c=0=d}$ are \emph{exact null geodesics}
\begin{equation}\label{specgeodcoor}
  \uRT=\uRT_\onscri=\text{constant}\comma\zRT=\zRT_\onscri=\text{constant}   \comma
\end{equation}
in the whole spacetime, with $\rRT$ being their affine parameter.
They approach  infinity $\scri$ along the
(double degenerate) principal null direction ${\kT=\cvil{\rRT}}$, which is
characterized by the parameter   ${\PSI=0}$, see Eq.~\eqref{cdPSIPHIrel} (or by ${\THT=0}$).
It can be shown that the Robinson-Trautman tetrad \eqref{RTtetrad} is parallelly
transported along these geodesics,
\begin{equation}
  \kT\ctr\covd\kT=0\comma\kT\ctr\covd\lT=0\comma\kT\ctr\covd\mT=0\commae
\end{equation}
and it is also invariant under a shift along the Killing vector $\cvil{\tKW}$,
i.e.,
${\lied_{\cv{\tKW}}\kT=0}$,
${\lied_{\cv{\tKW}}\lT=0}$,
${\lied_{\cv{\tKW}}\mT=0}$,
respecting thus a symmetry of the spacetime.
Consequently, we can naturally set the interpretation tetrad
${(\kI,\,\lI,\,\mI,\,\bI)} \equiv {(\kT,\,\lT,\,\mT,\,\bT)}$
in the \emph{whole} spacetime, not only asymptotically near $\scri$,
as in \eqref{genRTtetrad} for ${c=0}$, ${d=0}$.
As follows from Eqs.~\eqref{WeylRTTetr}, all components
of gravitational and electromagnetic fields are explicitly given by
\begin{equation}\label{WeylspTetr}
\begin{aligned}
  \WTP{\intT}{4} &=
  -\frac{3\accl^2}{\PRT}\,\Bigl(\mass +2\,\charge^2\accl\,\xKW - \frac{\charge^2}\rRT\Bigr)\,
  \,\frac1\rRT\commae\\
  \WTP{\intT}{3} &=
  \frac{3\accl}{\sqrt2\,\PRT}\,\Bigl(\mass + 2\,\charge^2\accl\,\xKW - \frac{\charge^2}\rRT\Bigr)\,
  \frac1{\rRT^2}\commae\\
  \WTP{\intT}{2} &=
  -\Bigl(\mass + 2\,\charge^2\accl\,\xKW - \frac{\charge^2}\rRT\Bigr)\,
  \frac1{\rRT^3}\comma
  \WTP{\intT}{1} = \WTP{\intT}{0} = 0\commae
\end{aligned}
\end{equation}
and
\begin{equation}\label{EMspTetr}
  \EMP{\intT}{2} = \frac{\charge\,\accl}{\sqrt2\,\PRT}\,\frac1\rRT\comma
  \EMP{\intT}{1} = -\frac{\charge}2\,\frac1{\rRT^2}\comma
  \EMP{\intT}{0} = 0\period
\end{equation}
Clearly, the leading terms in the ${1/\rRT}$ expansion give the previous general asymptotical
results \eqref{FieldsOnScri} with ${c=0=d}$.
In the case of anti--de~Sitter spacetime (${\mass=0}$, ${\charge=0}$) the field components
obviously identically vanish. In the general case
the fields have a radiative character (${\sim 1/\rRT}$) except
for a vanishing acceleration~$\accl$ and/or for ${\PRT=\infty}$.
The interesting \vague{static} limiting case ${\accl=0}$ has been already discussed
in Sec.~\ref{ssc:vanishingA}.
The case ${\PRT=\infty}$ corresponds to observers located at the privileged position where ${\GKW=0}$, i.e.,
on the axes ${\xKW=\xb}$ and ${\xKW=\xf}$ where the strings/struts are localized.
This is analogous to the situation when $\Lambda>0$ \cite{KrtousPodolsky:2003a},
and an electromagnetic field of accelerated test charges in flat spacetime: this is also not
radiative along the axis of symmetry, which is the direction of acceleration.

Let us also recall (see \cite{KrtousPodolsky:2003a}) that the affine parameter $\rRT$ coincides both with the
luminosity and the parallax distance for the congruence of the above null geodesics.
The radiative ${1/\rRT}$ fall-off of the fields is naturally measurable (even locally) by observers moving
radially to infinity, using both the luminosity and the parallax methods for determining the distance.

Concerning the other special family of ingoing null geodesics, ${c=0}$, ${d=-6/\Lambda}$, $\EPS=-1$,
it can be observed that the transformation given by \eqref{para}  becomes singular and
\eqref{genRTtetrad} is not thus justified. However, from Eqs.~\eqref{tangent}, \eqref{kPandTangVect},
and \eqref{RTtetrad}, with the condition \eqref{Hasympt} for fixing $\lI$, it follows that in this case
\begin{equation}\label{spectet}
\begin{aligned}
  \kI&\lteq-\frac{1}{\rafpc^2\,\afp^2}\biggl(\rRT^2\,\cv{\rRT}+\frac{6}{\Lambda}\,\cv{\uRT}\,\biggr)
     \lteq -\frac{6}{\Lambda}\frac1{\rafpc^2\afp^2}\,\lT \commae\\
 \lI&\lteq-\frac{\Lambda}{6}\rRT^2\,\cv{\rRT}
     \lteq -\frac{\Lambda}{6}\rafpc^2\afp^2\,\kT \commae
\end{aligned}
\end{equation}
which --- somewhat surprisingly --- fully agrees with expressions \eqref{genRTtetrad}
for the special case ${c=0}$, ${d=-6/\Lambda}$. Thus the interpretation tetrad along these geodesics is
equivalent to the tetrad \eqref{genRTtetrad} along the PND given by ${c=0}$, ${d=0}$
(which itself agrees with the Robinson-Trautman tetrad \eqref{RTtetrad}) after interchanging
${\kI\leftrightarrow\lI}$ and ${\mI\leftrightarrow-\bI}$, accompanied by a boost
\eqref{ae:boostrotation}
with ${B=-6/(\Lambda\rRT^2)}$. Components of the gravitational and electromagnetic fields can thus
easily be obtained from \eqref{WeylspTetr}, \eqref{EMspTetr}.
In particular, it follows that
${\WTP{\intT}{4} \lteq0\lteq \WTP{\intT}{3}}$, ${\WTP{\intT}{2}\sim \afp^{-3} }$,
${\WTP{\intT}{1}\sim \afp^{-4} }$, ${\WTP{\intT}{0}\sim \afp^{-5} }$, and
${\EMP{\intT}{2}\lteq 0}$, ${\EMP{\intT}{1}\sim \afp^{-2}}$, ${\EMP{\intT}{0}\sim \afp^{-3}}$.
Obviously, the radiative parts of the fields vanish along these special ingoing geodesics,
which agrees with the expression ${\PSI_\m=0}$ in \eqref{zeros2}, cf.\ \eqref{cdPSIPHIrel}.
Indeed, this direction is just the reflection of the PND $\kG_1$.

\section{Transformations of the components $\WTP{}{n}$ and $\EMP{}{n}$}
\label{apx:Transformations}

The Weyl tensor  can be parametrized by five standard complex coefficients defined as  components
with respect to a null tetrad (see, e.g., \cite{Krameretal:book}):
\begin{equation}\label{PsiDef}
\begin{aligned}
  \WTP{}{0} &= \spcm\WT_{\alpha\beta\gamma\delta}\,
    \kG^\alpha\,\mG^\beta\,\kG^\gamma\,\mG^\delta\commae\\
  \WTP{}{1} &= \spcm\WT_{\alpha\beta\gamma\delta}\,
    \kG^\alpha\,\lG^\beta\,\kG^\gamma\,\mG^\delta\commae\\
  \WTP{}{2} &= -\WT_{\alpha\beta\gamma\delta}\,
    \kG^\alpha\,\mG^\beta\,\lG^\gamma\,\bG^\delta\commae\\
  \WTP{}{3} &= \spcm\WT_{\alpha\beta\gamma\delta}\,
    \lG^\alpha\,\kG^\beta\,\lG^\gamma\,\bG^\delta\commae\\
  \WTP{}{4} &= \spcm\WT_{\alpha\beta\gamma\delta}\,
    \lG^\alpha\,\bG^\beta\,\lG^\gamma\,\bG^\delta\period
\end{aligned}
\end{equation}
Similarly, the tensor of electromagnetic field  is parametrized as
\begin{equation}\label{PhiDef}
\begin{aligned}
  \EMP{}{0} &= \EMF_{\alpha\beta}\,
    \kG^\alpha\,\mG^\beta\commae\\
  \EMP{}{1} &= {\textstyle\frac12}\,\EMF_{\alpha\beta}\,
    \bigl(\kG^\alpha\,\lG^\beta-\mG^\alpha\,\bG^\beta\bigr)\commae\\
  \EMP{}{2} &= \EMF_{\alpha\beta}\,
    \bG^\alpha\,\lG^\beta\period
\end{aligned}
\end{equation}
These transform in a well-known way under the following particular Lorentz
transformations. For a null rotation with $\kG$ fixed, $L\in\mathbb{C}$,
\begin{gather}
\begin{aligned}
  \kG &= \kO\commae\\
  \lG &= \lO + \bar L \mO + L \bO + L\bar L \kO\commae\\
  \mG &= \mO + L\kO\commae
\end{aligned}\label{ae:kfixed}\\
\notag\\
\begin{aligned}
  \WTP{}{0} &= \WTP{\refT}{0}\commae\\
  \WTP{}{1} &= {\bar L}\, \WTP{\refT}{0} + \WTP{\refT}{1} \commae\\
  \WTP{}{2} &= {\bar L}^2 \WTP{\refT}{0} + 2 {\bar L}\, \WTP{\refT}{1} + \WTP{\refT}{2}\commae\\
  \WTP{}{3} &= {\bar L}^3 \WTP{\refT}{0} + 3 {\bar L}^2 \WTP{\refT}{1} +
    3 {\bar L}\, \WTP{\refT}{2} + \WTP{\refT}{3}\commae\\
  \WTP{}{4} &= {\bar L}^4 \WTP{\refT}{0} + 4 {\bar L}^3 \WTP{\refT}{1} +
    6 {\bar L}^2 \WTP{\refT}{2} + 4 {\bar L}\, \WTP{\refT}{3} + \WTP{\refT}{4}\commae
\end{aligned}\label{ae:kfixedWeyl}\\
\notag\\
\begin{aligned}
  \EMP{}{0} &= \EMP{\refT}{0}\commae\\
  \EMP{}{1} &= {\bar L}\, \EMP{\refT}{0} + \EMP{\refT}{1} \commae\\
  \EMP{}{2} &= {\bar L}^2 \EMP{\refT}{0} + 2 {\bar L}\, \EMP{\refT}{1} + \EMP{\refT}{2}\period
\end{aligned}\label{ae:kfixedEM}
\end{gather}\pagebreak[1]
Under a null rotation with $\lG$ fixed, $K\in\mathbb{C}$,
\begin{gather}
\begin{aligned}
  \kG &= \kO + \bar K \mO + K \bO + K\bar K \lO\commae\\
  \lG &= \lO \commae\\
  \mG &= \mO + K\lO\commae
\end{aligned}\label{ae:lfixed}\\
\notag\\
\begin{aligned}
  \WTP{}{0} &= K^4 \WTP{\refT}{4} + 4 K^3 \WTP{\refT}{3} +
    6 K^2 \WTP{\refT}{2} + 4 K\, \WTP{\refT}{1} + \WTP{\refT}{0}\commae\\
  \WTP{}{1} &= K^3 \WTP{\refT}{4} + 3 K^2 \WTP{\refT}{3} +
    3 K\, \WTP{\refT}{2} + \WTP{\refT}{1}\commae\\
  \WTP{}{2} &= K^2 \WTP{\refT}{4} + 2 K\, \WTP{\refT}{3} + \WTP{\refT}{2}\commae\\
  \WTP{}{3} &= K\, \WTP{\refT}{4} + \WTP{\refT}{3} \commae\\
  \WTP{}{4} &= \WTP{\refT}{4}\commae
\end{aligned}\label{ae:lfixedWeyl}\\
\notag\\
\begin{aligned}
  \EMP{}{0} &= K^2 \EMP{\refT}{2} + 2 K\, \EMP{\refT}{1} + \EMP{\refT}{0}\commae\\
  \EMP{}{1} &= K\, \EMP{\refT}{2} + \EMP{\refT}{1} \commae\\
  \EMP{}{2} &= \EMP{\refT}{2}\period
\end{aligned}\label{ae:lfixedEM}
\end{gather}\\
Under a boost in the ${\kG\textdash\lG}$ plane and
a spatial rotation in the ${\mG\textdash\bG}$ plane given by
\begin{equation}\label{ae:boostrotation}
\begin{gathered}
  \kG = B\,\kO \comma   \lG = B^{-1}\, \lO \commae\\
  \mG = \exp(i\Phi)\,\mO \comma
\end{gathered}
\end{equation}
\mbox{}
$B, \Phi\in\mathbb{R}$, the components $\WTP{}{n}$ and $\EMP{}{n}$ transform as
\begin{gather}
\begin{aligned}
  \WTP{}{0} &= B^2\,\exp(2i\Phi)\; \WTP{\refT}{0} \commae\\
  \WTP{}{1} &= B\;\;\exp(i\Phi)\; \WTP{\refT}{1} \commae\\
  \WTP{}{2} &= \WTP{\refT}{2} \commae\\
  \WTP{}{3} &= B^{-1}\;\exp(-i\Phi)\; \WTP{\refT}{3} \commae\\
  \WTP{}{4} &= B^{-2}\,\exp(-2i\Phi)\; \WTP{\refT}{4}\commae
\end{aligned}\label{ae:boostrotationWeyl}\\
\notag\\
\begin{aligned}
  \EMP{}{0} &= B\;\;\exp(i\Phi)\; \EMP{\refT}{0} \commae\\
  \EMP{}{1} &= \EMP{\refT}{1} \commae\\
  \EMP{}{2} &= B^{-1}\,\exp(-i\Phi)\; \EMP{\refT}{2}\period
\end{aligned}\label{ae:boostrotationEMF}
\end{gather}

\end{document}